\documentclass[iop,numberedappendix,twocolappendix]{emulateapj}
\usepackage{apjfonts}
\usepackage{epsfig,color}
\usepackage{mathrsfs}
\usepackage{ulem}
\usepackage{amsmath}
\usepackage{empheq}
\usepackage{bm}

\def\bhm{M_{\bullet}}

\def\calS{{\cal S}}

\def\er{L_{\rm bol}/L_{\rm Edd}}
\def\ergs{\rm erg~s^{-1}}
\def\fblr{f_{_{\rm BLR}}}
\def\kms{\rm km~s^{-1}}

\def\mathdotM{\dot{\mathscr{M}}}

\def\Lhb{L_{_{\rm H\beta}}}

\def\pp{\prime\prime}
\def\pepsilon{\bar{\epsilon}}
\def\prblr{\langle\rblr\rangle}
\def\prblr{R_{_{{\rm H\beta},R-L}}} 
\def\rblr{R_{_{\rm H\beta}}}
\def\rblrl{\rblr-L_{5100}}
\def\rrhb{r_{_{\rm H\beta}}}

\def\Rg{R_{\rm g}}
\def\RFe{{\cal R}_{\rm Fe}}
\def\rhb{R_{_{\rm H\beta}}}

\def\sighb{\sigma_{\rm H\beta}}
\def\sunm{M_{\odot}}
\def\tauhb{\tau_{_{\rm H\beta}}}
\def\taurl{\tau_{_{R-L}}}

\def\Dhb{{\cal D}_{_{\rm H\beta}}}

\def\civ{C {\sc iv}}
\def\feii{Fe {\sc ii}}
\def\mgii{Mg~{\sc ii}}
\def\oiii{[O~{\sc iii}]}

\journalinfo{To appear in {\it The Astrophysical Journal}.}
\slugcomment{Received 2015 December 13; accepted 2016 April 20}

\begin{document}

\title{Supermassive Black Holes with High Accretion Rates in Active Galactic Nuclei. V.\\
A New Size-Luminosity Scaling Relation for the Broad-Line Region}

\author
{Pu Du\altaffilmark{1}, 
Kai-Xing Lu\altaffilmark{2,1},
Zhi-Xiang Zhang\altaffilmark{1},
Ying-Ke Huang\altaffilmark{1}, 
Kai Wang\altaffilmark{1}, 
Chen Hu\altaffilmark{1},
Jie Qiu\altaffilmark{1},
Yan-Rong Li\altaffilmark{1},\\
Xu-Liang Fan\altaffilmark{6},
Xiang-Er Fang\altaffilmark{9},
Jin-Ming Bai\altaffilmark{6}, 
Wei-Hao Bian\altaffilmark{8},
Ye-Fei Yuan\altaffilmark{9},\\
Luis C. Ho\altaffilmark{4,5} and
Jian-Min Wang\altaffilmark{1,3,*}\\
(SEAMBH collaboration)}

\altaffiltext{1}
{Key Laboratory for Particle Astrophysics, Institute of High Energy Physics,
Chinese Academy of Sciences, 19B Yuquan Road, Beijing 100049, China}

\altaffiltext{2}
{Astronomy Department, Beijing Normal University, Beijing 100875, China}

\altaffiltext{3}
{National Astronomical Observatories of China, Chinese Academy of Sciences,
 20A Datun Road, Beijing 100020, China}

\altaffiltext{4}
{Kavli Institute for Astronomy and Astrophysics, Peking University, Beijing 100871, China} 

\altaffiltext{5}
{Department of Astronomy, School of Physics, Peking University, Beijing 100871, China} 

\altaffiltext{6}{Yunnan Observatories, Chinese Academy of Sciences, Kunming 650011, China}

\altaffiltext{8}{Physics Department, Nanjing Normal University, Nanjing 210097, China}

\altaffiltext{9}{Department of Astronomy, University of Science and Technology of China, Hefei 
230026, China}

\altaffiltext{*}{Corresponding author; wangjm@ihep.ac.cn}

\begin{abstract} This paper reports results of the third-year campaign of
    monitoring super-Eddington accreting massive black holes (SEAMBHs) in
    active galactic nuclei (AGNs) between $2014-2015$. Ten new targets were
    selected from quasar sample of Sloan Digital Sky Survey (SDSS), which are
    generally more luminous than the SEAMBH candidates in last two years.
    H$\beta$ lags ($\tauhb$) in five of the 10 quasars have been successfully
    measured in this monitoring season. We find that the lags are generally
    shorter, by large factors, than those of objects with same optical
    luminosity, in light of the well-known $\rblrl$ relation. The five quasars
    have dimensionless accretion rates of $\mathdotM=10-10^3$. Combining
    measurements of the previous SEAMBHs, we find that the reduction of
    H$\beta$ lags tightly depends on accretion rates,
    $\tauhb/\taurl\propto\mathdotM^{-0.42}$, where $\taurl$ is the H$\beta$ lag
    from the normal $\rblrl$ relation. Fitting 63 mapped AGNs, we present a new
    scaling relation for the broad-line region:
    $\rhb=\alpha_1\ell_{44}^{\beta_1}\,\min\left[1,\left(\mathdotM/\mathdotM_c\right)^{-\gamma_1}\right]$,
    where $\ell_{44}=L_{5100}/10^{44}\,\ergs$ is 5100 \AA\, continuum
    luminosity, and coefficients of $\alpha_1=(29.6_{-2.8}^{+2.7})$\,lt-d,
    $\beta_1=0.56_{-0.03}^{+0.03}$, $\gamma_1=0.52_{-0.16}^{+0.33}$ and
$\mathdotM_c=11.19_{-6.22}^{+2.29}$. This relation is applicable to AGNs
    over a wide range of accretion rates, from $10^{-3}$ to $10^3$.
    Implications of this new relation are briefly discussed.  \end{abstract}


\section{Introduction}
This is the fifth paper of the series reporting the ongoing large campaign of monitoring 
Super-Eddington Accreting Massive Black Holes (SEAMBHs) in active galaxies and quasars 
starting from October 2012. One of the major goals of the campaign is to search for massive 
black holes with extreme accretion rates through reverberation mapping (RM) of broad emission lines 
and continuum. Results from the campaigns in 2012--2013 and 2013--2014 have been reported by 
Du et al. (2014, 2015, hereafter Papers I and IV), Wang et al. (2014, Paper II) and Hu et al. (2015, 
Paper III). This paper carries out the results of SEAMBH2014 sample, which was monitored from 
September 2014 to June 2015. With the three monitoring years of observations,  
we build up a new scaling relation of the broad-line region (BLR) in this paper.

\begin{deluxetable*}{lllllccrr}[!ht]
\tablecolumns{9}
\tablewidth{\textwidth}
\setlength{\tabcolsep}{4pt}
\tablecaption{The SEAMBH project: observational results}
\tabletypesize{\scriptsize}
\tablehead{
\colhead{Object}                      &
\colhead{$\alpha_{2000}$}             &
\colhead{$\delta_{2000}$}             &
\colhead{redshift}                    &
\colhead{monitoring period}           &
\colhead{$N_{\rm spec}$}              &
\colhead{}                            &
\multicolumn{2}{c}{Comparison stars}  \\ \cline{8-9}
\colhead{}                            &
\colhead{}                            &
\colhead{}                            &
\colhead{}                            &
\colhead{}                            &
\colhead{}                            &
\colhead{}                            &
\colhead{$R_*$}                       &
\colhead{P.A.}                  
}
\startdata
\multicolumn{9}{c}{First phase: SEAMBH2012 sample}\\ \hline 
Mrk 335                  & 00 06 19.5 & $+$20 12 10 & 0.0258 & Oct., 2012 $-$ Feb., 2013 & 91 & & $ 80^{\pp}.7$ & $ 174.5^{\circ}$ \\
Mrk 1044                 & 02 30 05.5 & $-$08 59 53 & 0.0165 & Oct., 2012 $-$ Feb., 2013 & 77 & & $207^{\pp}.0$ & $-143.0^{\circ}$ \\
IRAS 04416+1215          & 04 44 28.8 & $+$12 21 12 & 0.0889 & Oct., 2012 $-$ Mar., 2013 & 92 & & $137^{\pp}.9$ & $ -55.0^{\circ}$ \\
Mrk 382                  & 07 55 25.3 & $+$39 11 10 & 0.0337 & Oct., 2012 $-$ May., 2013 &123 & & $198^{\pp}.4$ & $ -24.6^{\circ}$ \\
Mrk 142                  & 10 25 31.3 & $+$51 40 35 & 0.0449 & Nov., 2012 $-$ Apr., 2013 &119 & & $113^{\pp}.1$ & $ 155.2^{\circ}$ \\
MCG $+06-26-012$         & 11 39 13.9 & $+$33 55 51 & 0.0328 & Jan., 2013 $-$ Jun., 2013 & 34 & & $204^{\pp}.3$ & $  46.1^{\circ}$ \\
IRAS F12397+3333         & 12 42 10.6 & $+$33 17 03 & 0.0435 & Jan., 2013 $-$ May., 2013 & 51 & & $189^{\pp}.0$ & $ 130.0^{\circ}$ \\
Mrk 486                  & 15 36 38.3 & $+$54 33 33 & 0.0389 & Mar., 2013 $-$ Jul., 2013 & 45 & & $193^{\pp}.8$ & $-167.0^{\circ}$ \\
Mrk 493                  & 15 59 09.6 & $+$35 01 47 & 0.0313 & Apr., 2013 $-$ Jun., 2013 & 27 & & $155^{\pp}.3$ & $  98.5^{\circ}$ \\ \hline
\multicolumn{9}{c}{Second phase: SEAMBH2013 sample}\\ \hline
SDSS J075101.42+291419.1 & 07 51 01.4 & $+$29 14 19 & 0.1208 & Nov., 2013 $-$ May., 2014 & 38 & & $133^{\pp}.3$ & $ -41.3^{\circ}$ \\
SDSS J080101.41+184840.7 & 08 01 01.4 & $+$18 48 40 & 0.1396 & Nov., 2013 $-$ Apr., 2014 & 34 & & $118^{\pp}.8$ & $ -98.2^{\circ}$ \\
SDSS J080131.58+354436.4 & 08 01 31.6 & $+$35 44 36 & 0.1786 & Nov., 2013 $-$ Apr., 2014 & 31 & & $100^{\pp}.0$ & $ 145.2^{\circ}$ \\
SDSS J081441.91+212918.5 & 08 14 41.9 & $+$21 29 19 & 0.1628 & Nov., 2013 $-$ May., 2014 & 34 & & $ 79^{\pp}.0$ & $  73.9^{\circ}$ \\
SDSS J081456.10+532533.5 & 08 14 56.1 & $+$53 25 34 & 0.1197 & Nov., 2013 $-$ Apr., 2014 & 27 & & $164^{\pp}.5$ & $-172.9^{\circ}$ \\
SDSS J093922.89+370943.9 & 09 39 22.9 & $+$37 09 44 & 0.1859 & Nov., 2013 $-$ Jun., 2014 & 26 & & $175^{\pp}.1$ & $-139.0^{\circ}$ \\ \hline
\multicolumn{9}{c}{Third phase: SEAMBH2014 sample}\\ \hline
SDSS J075949.54+320023.8 & 07 59 49.5 & $+$32 00 24 & 0.1880 & Sep., 2014 $-$ May., 2015 & 27 & & $109^{\pp}.2$ & $ -48.3^{\circ}$ \\
SDSS J080131.58+354436.4 & 08 01 31.6 & $+$35 44 36 & 0.1786 & Oct., 2014 $-$ May., 2015 & 19 & & $139^{\pp}.2$ & $ -85.3^{\circ}$ \\
SDSS J084533.28+474934.5 & 08 45 33.3 & $+$47 49 35 & 0.3018 & Sep., 2014 $-$ Apr., 2015 & 18 & & $205^{\pp}.5$ & $-126.4^{\circ}$ \\
SDSS J085946.35+274534.8 & 08 59 46.4 & $+$27 45 35 & 0.2438 & Sep., 2014 $-$ Jun., 2015 & 26 & & $169^{\pp}.8$ & $ -89.1^{\circ}$ \\
SDSS J102339.64+523349.6 & 10 23 39.6 & $+$52 33 50 & 0.1364 & Oct., 2014 $-$ Jun., 2015 & 26 & & $123^{\pp}.2$ & $ 108.1^{\circ}$ 
\enddata
\tablecomments{\footnotesize
This table follows the contents Table 1 in Paper IV. We denote the samples monitored during the 
2012--2013, 2013--2014 and 2014-2015 observing seasons as SEAMBH2012, SEAMBH2013 and SEAMBH2014, 
respectively. $N_{\rm spec}$ is the numbers of spectroscopic epochs, $R_*$ is the angular distance
between the object and the comparison star and PA the position angle from the AGN to the comparison 
star. We marked the time lag of J080131 as ``uncertain" in PaperIV, however we pick it up here because 
its lag reported in Paper IV is highly consistent with the number measured in the present paper.
}
\end{deluxetable*}
\vspace{1.0cm}

Reverberation mapping (RM) technique, measuring the delayed echoes of broad lines to the varying
ionizing continuum (Bahcall et al. 1972; Blandford \& McKee 1982; Peterson 1993), is a powerful 
tool to probe the kinematics and geometry of the BLRs in the time domain.
Countless clouds, which contribute to the smooth profiles 
of the broad emission lines (e.g., Arav et al. 1997), are distributed in the vicinity of supermassive 
black hole (SMBH), composing the BLR. As an observational consequence 
of photonionization powered by the accretion disk under the deep gravitational potential of the SMBH, 
the profiles of the lines are broadened, and line emission from the clouds reverberate in response 
to the varying ionizing continuum. The reverberation is delayed because of light travel difference 
between H$\beta$ and ionizing photons and is thus expected to deliver 
information on the kinematics and structure of the BLR. The unambiguous reverberation 
of the lines, detected by monitoring campaigns from ultraviolet to optical 
bands since the late 1980s, supports this picture of the central engine of AGNs
(e.g., Clavel et al. 1991; Peterson et al. 1991, 1993; Maoz 
et al. 1991; Wanders et al. 1993; Dietrich et al. 1993, 1998, 2012; Kaspi et al. 2000; Denney et al. 
2006, 2010; Bentz et al. 2009, 2014; Grier et al. 2012; Papers I-IV;
Barth et al. 2013, 2015; Shen et al. 2015a,b). The $\rblrl$ relation was first discussed 
by Koratkar \& Gaskell (1991) and Peterson (1993). 
Robust RM results for 41 AGNs in the last four decades 
lead to a simple, highly significant correlation of the form
\begin{equation}
    \rblr\approx \alpha_0\,\ell_{44}^{\beta_0},
\end{equation}
where $\ell_{44}=L_{5100}/10^{44}\ergs$ is the 5100 \AA\, luminosity in units of $10^{44}\ergs$ 
(corrected for host galaxy contamination)
and $\rblr=c\tauhb$ is the emissivity-weighted radius of the BLR
(Kaspi et al. 2000; Bentz et al. 2013). We refer to this type of correlation as the normal $\rblrl$ 
relationship. The constants $\alpha_0$ and $\beta_0$ 
differ slightly from one study to the next, depending on the number of sources and their exact 
luminosity range (e.g., Kilerci Eser et al. 2015). For sub-Eddington accreting AGNs,
$\alpha_0=35.5$ ltd and $\beta_0=0.53$, but for SEAMBHs they are different (see Paper IV). 

As reported in Paper IV, some objects from the SEAMBH2012 and SEAMBH2013 samples have much shorter 
H$\beta$ lags compared with objects with similar luminosity, and the $\rblrl$ relation has a large 
scatter if they are included. In particular, the reduction of the lags increases with the dimensionless 
accretion rate, defined as
$\mathdotM=\dot{M}_{\bullet}/L_{\rm Edd}c^{-2}$, where $\dot{M}_{\bullet}$ is the accretion rate, 
$L_{\rm Edd}$ is the Eddington luminosity and $c$ is the speed of light. 
Furthermore, it has been found, so far in the present campaigns, that SEAMBHs have a range of accretion 
rates from a few to $\sim 10^3$. This kind of shortened H$\beta$ lags was discovered in 
the current SEAMBH project (a comparison with previous campaigns is given in Section 6.5). 
Such high accretion rates are characteristic
of the regime of slim accretion disks (Abramowicz et al. 
1988; Szuszkiewicz et al. 1996; Wang \& Zhou 1999; Wang et al. 1999; Mineshige et al. 2000; Wang \& 
Netzer 2003; Sadowski 2009). These interesting properties needed to be confirmed with observations.  
We aim to explore whether we can define a new scaling relation, 
$\rhb=\rhb(L_{5100},\mathdotM)$, which links the size of the BLR to both the
AGN luminosity {\it and}\ accretion rate.

We report new results from SEAMBH2014.  We describe target selection, observation details and data 
reduction in \S2. H$\beta$ lags, BH 
mass and accretion rates are provided in \S3. Properties of H$\beta$ lags are discussed in \S4,
and a new scaling relation of H$\beta$ lags is established in \S5. Section 6 introduces 
the fundamental plane, which is used to estimate accretion rates from single-epoch spectra,
for application of the new size-luminosity scaling relation of the BLR. 
Brief discussions of the shortened lags are presented in \S7. We draw conclusions in \S8. 
Throughout this work we assume a standard 
$\Lambda$CDM cosmology with $H_0=67~{\rm km~s^{-1}~Mpc^{-1}}$, $\Omega_{\Lambda}=0.68$ and 
$\Omega_{\rm M}=0.32$ (Ade et al. 2014).

\begin{deluxetable*}{rlcrllcrlcrll}
  \tablecolumns{13}
  \setlength{\tabcolsep}{3pt}
  \tablewidth{0pc}
  \tablecaption{Light curves of J075949 and J080131}
  \tabletypesize{\scriptsize}
  \tablehead{
      \multicolumn{6}{c}{J075949}        &
      \colhead{}                         &
      \multicolumn{6}{c}{J080131}        \\ \cline{1-6}\cline{8-13}
      \multicolumn{2}{c}{Photometry}     &
      \colhead{}                         &
      \multicolumn{3}{c}{Spectra}        &
      \colhead{}                         &
      \multicolumn{2}{c}{Photometry}     &
      \colhead{}                         &
      \multicolumn{3}{c}{Spectra}        \\ \cline{1-2}\cline{4-6}\cline{8-9}\cline{11-13}
      \colhead{JD}                       &
      \colhead{mag}                      &
      \colhead{}                         &
      \colhead{JD}                       &
      \colhead{$F_{5100}$}               &
      \colhead{$F_{\rm H\beta}$}         &
      \colhead{}                         &
      \colhead{JD}                       &
      \colhead{mag}                      &
      \colhead{}                         &
      \colhead{JD}                       &
      \colhead{$F_{5100}$}               &
      \colhead{$F_{\rm H\beta}$}
         }
\startdata
  29.374 & $17.373\pm 0.007$  & &   76.365 & $ 2.773\pm 0.029$ & $ 2.469\pm 0.038$ & &   60.324 & $17.757\pm 0.010$  & &  112.294 & $ 2.096\pm 0.016$ & $ 0.853\pm 0.027$ \\
  30.360 & $17.375\pm 0.008$  & &   80.319 & $ 2.864\pm 0.017$ & $ 2.461\pm 0.031$ & &   62.314 & $17.720\pm 0.010$  & &  116.394 & $ 2.076\pm 0.016$ & $ 0.893\pm 0.029$ \\
  32.352 & $17.406\pm 0.009$  & &   83.314 & $ 2.724\pm 0.038$ & $ 2.366\pm 0.041$ & &   63.299 & $17.734\pm 0.010$  & &  119.336 & $ 2.106\pm 0.021$ & $ 0.831\pm 0.031$ \\
  33.339 & $17.416\pm 0.009$  & &   86.422 & $ 2.795\pm 0.011$ & $ 2.535\pm 0.023$ & &   68.397 & $17.721\pm 0.013$  & &  135.324 & $ 2.086\pm 0.026$ & $ 0.807\pm 0.033$ \\
  34.331 & $17.408\pm 0.009$  & &   89.378 & $ 2.888\pm 0.012$ & $ 2.402\pm 0.032$ & &   77.318 & $17.746\pm 0.009$  & &  139.319 & $ 2.074\pm 0.028$ & $ 0.862\pm 0.035$ \\
  \enddata
  \tablecomments{\footnotesize
      JD: Julian dates from 2,456,900; $F_{5100}$ and $F_{\rm H\beta}$ are fluxes at $(1+z)5100$ \AA\, 
      and H$\beta$ emission lines in units of $10^{-16}{\rm erg~s^{-1}~cm^{-2}~ \AA^{-1}}$ and 
      $10^{-14}{\rm erg~s^{-1}~cm^{-2}}$.
      (This table is available in its entirety in a machine-readable form in the online journal. A portion is shown here for guidance regarding its form and content.)
      }
\end{deluxetable*}

\section{Observations and data reduction}
\subsection{Target Selection}
We followed the procedures for selecting SEAMBH candidates described in Paper IV.
We used the fitting procedures to measure H$\beta$ profile and 5100 \AA\,
luminosity of SDSS quasar spectra described by Hu et al. 
(2008a,b). Following the standard assumption that the BLR gas is virialized, 
we estimate the BH mass as
\begin{equation}
\bhm=\fblr \frac{\rblr V_{\rm FWHM}^2}{G}=1.95\times 10^6~\fblr V_{3}^2\tau_{10}~\sunm,
\end{equation}
where $\rblr=c\tauhb$, $\tauhb$ is the H$\beta$ lag measured in the rest frame, 
$\tau_{10}=\tauhb/10$days, $G$ is the gravitational constant, and 
$V_{3}=V_{\rm FWHM}/10^3\kms$ is the 
full-width-half-maximum (FWHM) of the H$\beta$ line profile in units of 
$10^3\kms$. We take the virial factor $\fblr=1$ in our series of papers
(see some discussions in Paper IV). 

In order to select AGNs with high accretion rates, we
employed the formulation of accretion rates derived from the standard disk model of Shakura \& 
Sunyaev (1973). In the standard model it is assumed that the disk gas is rotating with
Keplerian angular momentum, and 
thermal equilibrium is localized between viscous dissipation and blackbody cooling. 
Observationally, this model is supported from fits of the so-called big blue bump in quasars
(Czerny \& Elvis 1987; Wandel \& Petrosian 1988; Sun \& Malkan 1989; Laor \& Netzer 1989; Collin 
et al. 2002; Brocksopp et al. 2006; Kishimoto et al. 2008; Davis \& Laor 2011; Capellupo et al. 
2015). The dimensionless accretion rate is given by
\begin{equation}
\mathdotM=20.1\left(\frac{\ell_{44}}{\cos i}\right)^{3/2}m_7^{-2},
\label{eq:SS_2}
\end{equation}
where $m_7=\bhm/10^7\sunm$ (see Papers II and IV) and $i$ is the inclination angle to the line 
of sight of the disk. We take $\cos i=0.75$, which represents a mean disk inclination for a type 
1 AGNs with a torus covering factor of about 0.5 (it is assumed that the torus axis is co-aligned 
with the disk axis). Previous studies estimate $i\approx0-45^{\circ}$ [e.g.,  
Fischer et al. (2014) find a inclination range of $i\approx10^{\circ}-45^{\circ}$, whereas 
Pancoast et al. (2014) quote
$i\approx 5^{\circ}-45^{\circ}$; see also supplementary materials in Shen \& Ho (2014)], 
which results in $\Delta \log \mathdotM=1.5\Delta \log \cos i \lesssim 0.15$ from Equation (3). This 
uncertainty is significantly smaller than the average uncertainty on $\mathdotM$ ($\sim 0.3-0.5$ dex) 
in the present paper, and is thus ignored.
Equation (3) applies to AGNs that have accretion rates 
$10^{-2}\lesssim\mathdotM\lesssim 3\times 10^3$, namely excluding the regimes of 
advection-dominated accretion flows (ADAF; Narayan \& Yi 1994) and of flows with hyperaccretion rates 
($\mathdotM\ge 3\times 10^3$; see Appendix A for the validity of Equation 3 for SEAMBHs).

Using the normal $\rblrl$ relation (Bentz et al. 2013), we fitted all the quasar spectra 
in SDSS Data Release 7 by the procedures in Hu et al. (2008a, b) and  
applied Equations (2) and (3) to select high$-\mathdotM$ targets. 
We ranked quasars in terms of $\mathdotM$ and chose ones as candidates 
with the highest $\mathdotM$. We found that the high$-\mathdotM$ quasars are characterized by
1) strong optical \feii\ lines; 
2) relatively narrow H$\beta$ lines ($\lesssim 2000\kms$);
3) weak \oiii\ lines; and 4) steep 2--10 keV spectra (Wang et al. 2004).
These properties are similar to those of NLS1s (Osterbrock \& Pogge 
1987; Boroson \& Green 1992), but most of the candidates have more extreme accretion rates 
(a detailed comparison of SEAMBH properties with normal quasars will be carried out in a 
separate paper). Considering that the lags of all targets should be measured 
within one observing season, and taking into consideration the limitations of the  weather of 
the Lijiang Station of Yunnan Observatory (periods between June and September are raining seasons 
there), we only chose objects with maximum estimated lags of about 100 days or so (the monitoring periods 
should be at least a few times the presumed lags). Also, to ensure adequate signal-to-noise ratio (S/N) 
for measurements of light curves, we restricted the targets to a redshift range 
of $z=0.1-0.3$ and magnitudes $r^{\prime}\le 18.0$.
The fraction of radio-loud objects with $\mathdotM>3$
is not high. We discarded radio-loud objects\footnote{It has been realised that high-accretion 
rate AGNs are usually radio-quiet (Greene \& Ho 2006), although there are a few NLS1s
reported to be radio-loud. The fraction of radio-loud AGNs decreases with increasing accretion 
rate (Ho 2002, 2008).} based on available FIRST observations, in order 
to avoid H$\beta$ reverberations potentially affected by nonthermal emission from relativistic 
jets, or optical continuum emission strongly contaminated by jets. We chose about 20 targets
for photometry monitoring, which served as a preselection to trigger follow-up
spectroscopic monitoring. The photometric monitoring yielded 10 targets with
significant variations ($\gtrsim 0.1$ magnitudes), and time
lags were successfully measured for 5 objects (Table 1). For an overview of our
entire ongoing campaign, Table 1 also lists samples from SEAMBH2012 and
SEAMBH2013.

To summarize: we have selected about 30 targets for spectroscopic monitoring during the last 
three years (2012--2014). The successful rate of the monitoring project is about 2/3. Our 
failure to detect a lag for the remaining 1/3 of the sample are either due to low-amplitude 
variability or bad weather that leads to poor monitoring cadence. In particular, the SEAMBH2014 
observations were seriously affected by the El Ni\~no phenomenon.

\begin{deluxetable*}{rlcrllcrlcrll}[!h]
  \tablecolumns{13}
  \setlength{\tabcolsep}{3pt}
  \tablewidth{0pc}
  \tablecaption{Light curves of J084533 and J085946}
  \tabletypesize{\scriptsize}
  \tablehead{
      \multicolumn{6}{c}{J084533}        &
      \colhead{}                         &
      \multicolumn{6}{c}{J085946}        \\ \cline{1-6}\cline{8-13}
      \multicolumn{2}{c}{Photometry}     &
      \colhead{}                         &
      \multicolumn{3}{c}{Spectra}        &
      \colhead{}                         &
      \multicolumn{2}{c}{Photometry}     &
      \colhead{}                         &
      \multicolumn{3}{c}{Spectra}        \\ \cline{1-2}\cline{4-6}\cline{8-9}\cline{11-13}
      \colhead{JD}                       &
      \colhead{mag}                      &
      \colhead{}                         &
      \colhead{JD}                       &
      \colhead{$F_{5100}$}               &
      \colhead{$F_{\rm H\beta}$}         &
      \colhead{}                         &
      \colhead{JD}                       &
      \colhead{mag}                      &
      \colhead{}                         &
      \colhead{JD}                       &
      \colhead{$F_{5100}$}               &
      \colhead{$F_{\rm H\beta}$}
         }
\startdata
  29.417 & $17.803\pm 0.008$  & &   75.402 & $ 1.728\pm 0.026$ & $ 1.145\pm 0.047$ & &   30.428 & $17.401\pm 0.008$  & &   90.382 & $ 2.651\pm 0.015$ & $ 1.693\pm 0.024$ \\
  30.398 & $17.797\pm 0.008$  & &   80.395 & $ 1.721\pm 0.019$ & $ 1.085\pm 0.033$ & &   33.403 & $17.399\pm 0.007$  & &   97.433 & $ 2.593\pm 0.033$ & $ 1.728\pm 0.038$ \\
  33.376 & $17.777\pm 0.009$  & &   84.365 & $ 1.680\pm 0.020$ & $ 1.046\pm 0.035$ & &   36.410 & $17.406\pm 0.005$  & &  104.293 & $ 2.392\pm 0.024$ & $ 1.835\pm 0.034$ \\
  36.368 & $17.780\pm 0.009$  & &   91.326 & $ 1.673\pm 0.016$ & $ 1.086\pm 0.028$ & &   48.357 & $17.422\pm 0.008$  & &  111.261 & $ 2.583\pm 0.022$ & $ 1.595\pm 0.037$ \\
  38.411 & $17.816\pm 0.017$  & &  104.409 & $ 1.681\pm 0.015$ & $ 1.069\pm 0.029$ & &   51.353 & $17.440\pm 0.006$  & &  116.448 & $ 2.512\pm 0.025$ & $ 1.697\pm 0.027$ \\
  \enddata
  \tablecomments{\footnotesize
      This table is available in its entirety in a machine-readable form in the online journal. A portion is shown here for guidance regarding its form and content.
      }
\end{deluxetable*}

\subsection{Photometry and Spectroscopy}
The SEAMBH project uses the Lijiang 2.4m telescope, which has an alt-azimuth
Ritchey-Chr\'etien mount with a field de-rotator that enables two objects to be
positioned along the same long slit. It is located in Lijiang and is operated
by Yunnan Observatories. We adopted the same observational procedures described
in detail in Paper I, which also introduces the telescope and spectrograph. We
employed the Yunnan Faint Object Spectrograph and Camera (YFOSC), which has a
back-illuminated 2048$\times$4608 pixel CCD covering a field of
$10^{\prime}\times 10^{\prime}$. During the spectroscopic observation,
we put the target and a nearby comparison star 
into the slit simultaneously, which can provide
high-precision flux calibration. As in SEAMBH2013 (Paper IV), we adopted a
$5^{\pp}$-wide slit to minimize the influence of atmospheric differential
refraction, and used Grism 3 with a spectral resolution of 2.9 \AA/pixel and
wavelength coverage of 3800--9000 \AA.  To check the accuracy of 
spectroscopic calibration, we performed differential photometry of the targets
using some other stars in the same field. We used an SDSS $r^{\prime}$-band filter
for photometry to avoid the potential contamination by emission lines such as
H$\beta$ and H$\alpha$. Photometric and spectroscopic exposure times are
typically 10 and 60 min, respectively.

The reduction of the photometry data was done in a standard way using 
{\tt IRAF} routines. Photometric light curves were 
produced by comparing the instrumental magnitudes to those of standard stars in the field (see, 
e.g., Netzer et al. 1996, for details). 
The radius for the aperture photometry is typically $\sim4^{\pp}$ (seeing $\sim1.5^{\pp}-2^{\pp}$), 
and background is determined from an annulus with radius $8^{\pp}.5$ to $17^{\pp}$.
The uncertainties on the photometric measurements include 
the fluctuations due to photon statistics and the scatter in the measurement of the stars used. 

The spectroscopic data were also reduced with IRAF. The extraction width is fixed to $8^{\pp}.5$,
and the sky regions are set to $7^{\pp}.4-14^{\pp}.1$ on both sides of the extracted region. 
The average S/N of the 5100 \AA\ continuum of individual spectra are from $\sim$16 to 
$\sim$22, except for J085946, which only has S/N $\approx $12.
The flux of spectroscopic data was calibrated by simultaneously observing a nearby comparsion star 
along the slit (see Paper I). The fiducial spectra of the comparison stars are generated 
using observations from several nights with the best weather conditions. 
The absolute fluxes of the fiducial spectra are calibrated using additional
spectrophotometric standard stars observed in those nights.  Then, the in-slit comparison stars are used 
as standards to calibrate the spectra of targets observed in each night. The sensitivity as a function of 
wavelength is produced by comparing the observed spectrum of the comparison star to its fiducial spectrum. 
Finally, the sensitivity function is applied to calibrate the observed AGN spectrum\footnote{The 
uncertainty of our absolute flux calibration is $\lesssim$10\%. We multiply the fiducial spectra of the 
in-slit stars with the bandpass of the SDSS r$^{\prime}$ filter and compare their synthesized  magnitudes 
with the magnitudes found in the SDSS database. The maximum difference is $\lesssim$10\%. }.
The procedures adopted here resemble the method used by Maoz et al. (1990) and Kaspi et al. (2000). 
In order to illustrate the invariance of the comparison stars, we show the light curves from the differential 
photometry of the comparison stars in Appendix B. It is clear that their fluxes are very stable
and the variations are less than $\sim$1\%.

The calibration method of van Groningen \& Wanders (1992), based on the \oiii\ emission line and 
popularily used in many RM campaigns (e.g., Peterson et al. 1998; Bentz et al.
2009; Denney et al. 2010; Grier et al. 2012; Barth et al. 2015), 
is not suitable for SEAMBHs.  \oiii$\lambda5007$\ tends to be weak in SEAMBHs 
(especially for the objects in SEAMBH2013-2014), and, even worse, is blended with strong \feii\ around 5016 \AA.
Applying the calibration method to SEAMBHs results in large statistical
(caused by the weakness of \oiii) and systematic (caused by the variability of
\feii; see Paper III) uncertainties. The method based on in-slit comparison stars, 
used in our campaign, does not rely on \oiii\ and provides accurate flux calibration
for the spectra of SEAMBHs. For comparison, in Paper I we measured the \oiii\ 
fluxes in the calibrated spectra of three objects in SEAMBH2012 with relatively 
strong \oiii; the variation of their \oiii\ flux is on the order of $\sim$3\%.  
This clearly demonstrates the robustness of our flux calibration method based 
on in-slit comparison stars.

The procedures to measure the 5100 \AA\ and H$\beta$
flux are nearly the same as those given in Paper I. The continuum
beneath H$\beta$ line is determined by interpolation of two nearby bands 
(4740--4790 \AA\ and 5075 -5125 \AA) in the rest frame. These two bands have 
minimal contamination from emission lines. The flux of H$\beta$ is measured by 
integrating the band between 4810 and 4910 \AA\ after subtraction of the continuum; 
the H$\beta$ band is chosen to avoid the influence from \feii\ lines. 
The 5100 \AA\ flux is taken to be the median over the region 5075-5125 \AA. 
Detailed information of the
observations is provided in Table 1. All the photometry and continuum and
H$\beta$ light curves for the five objects with successfully detected lags are
listed in Tables $2-4$ and shown in Figure 1. We also calculated the mean and
RMS (root mean square) spectra and present them in Appendix C.

\begin{figure*}[t!]
\begin{center}
\includegraphics[angle=0,width=0.48\textwidth]{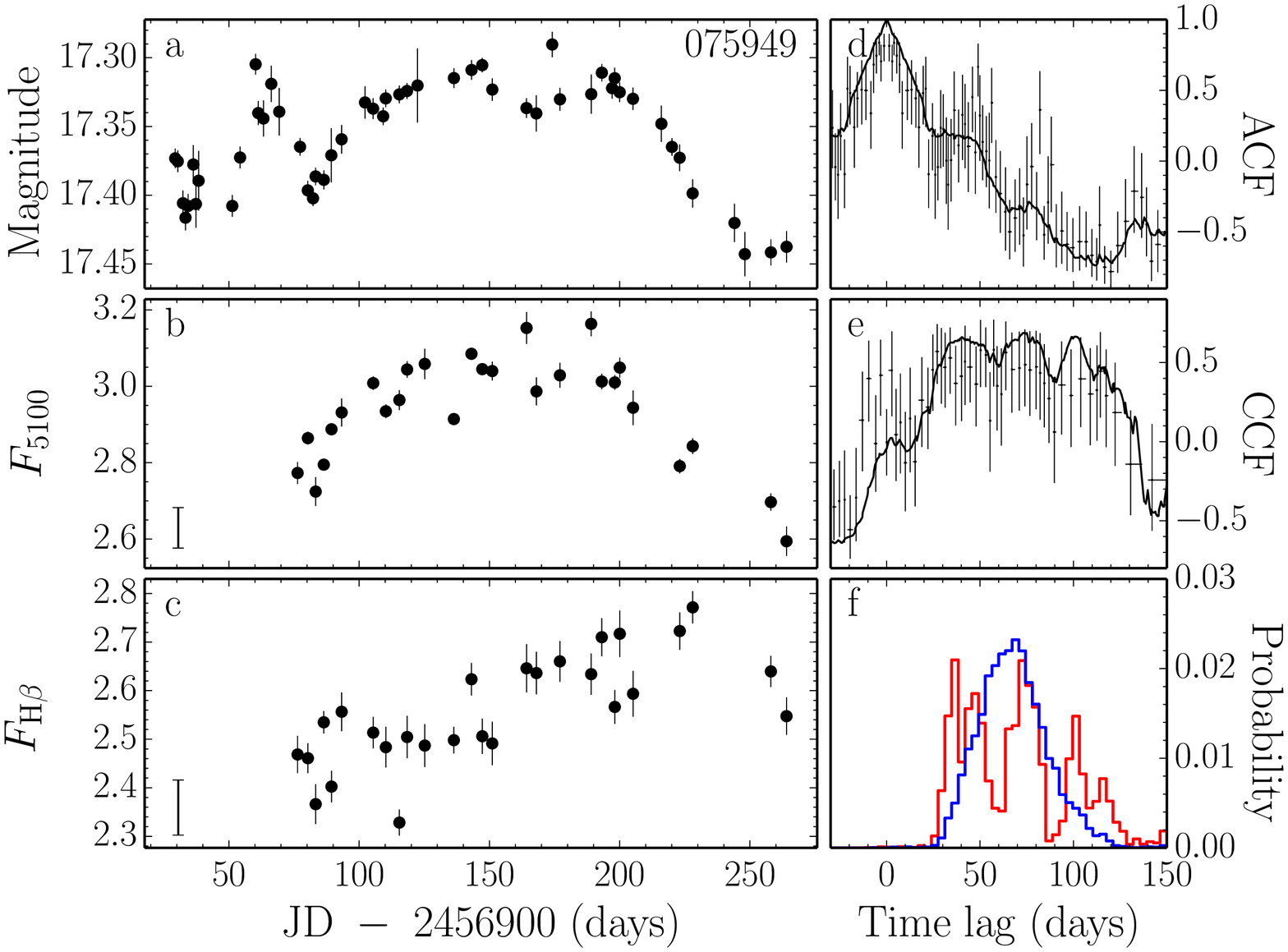}\hspace{0.5cm}
\includegraphics[angle=0,width=0.48\textwidth]{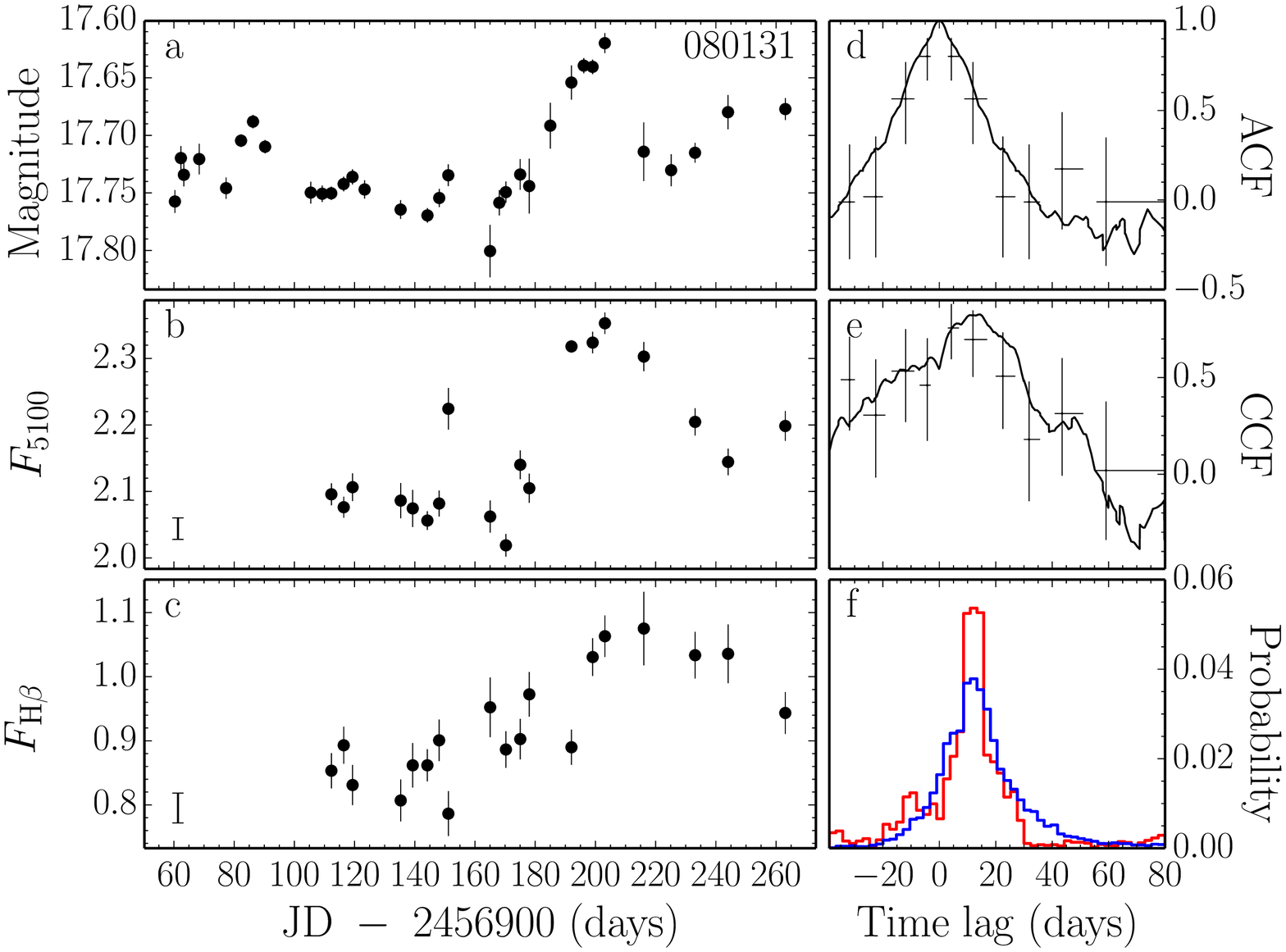} \\ 
\includegraphics[angle=0,width=0.48\textwidth]{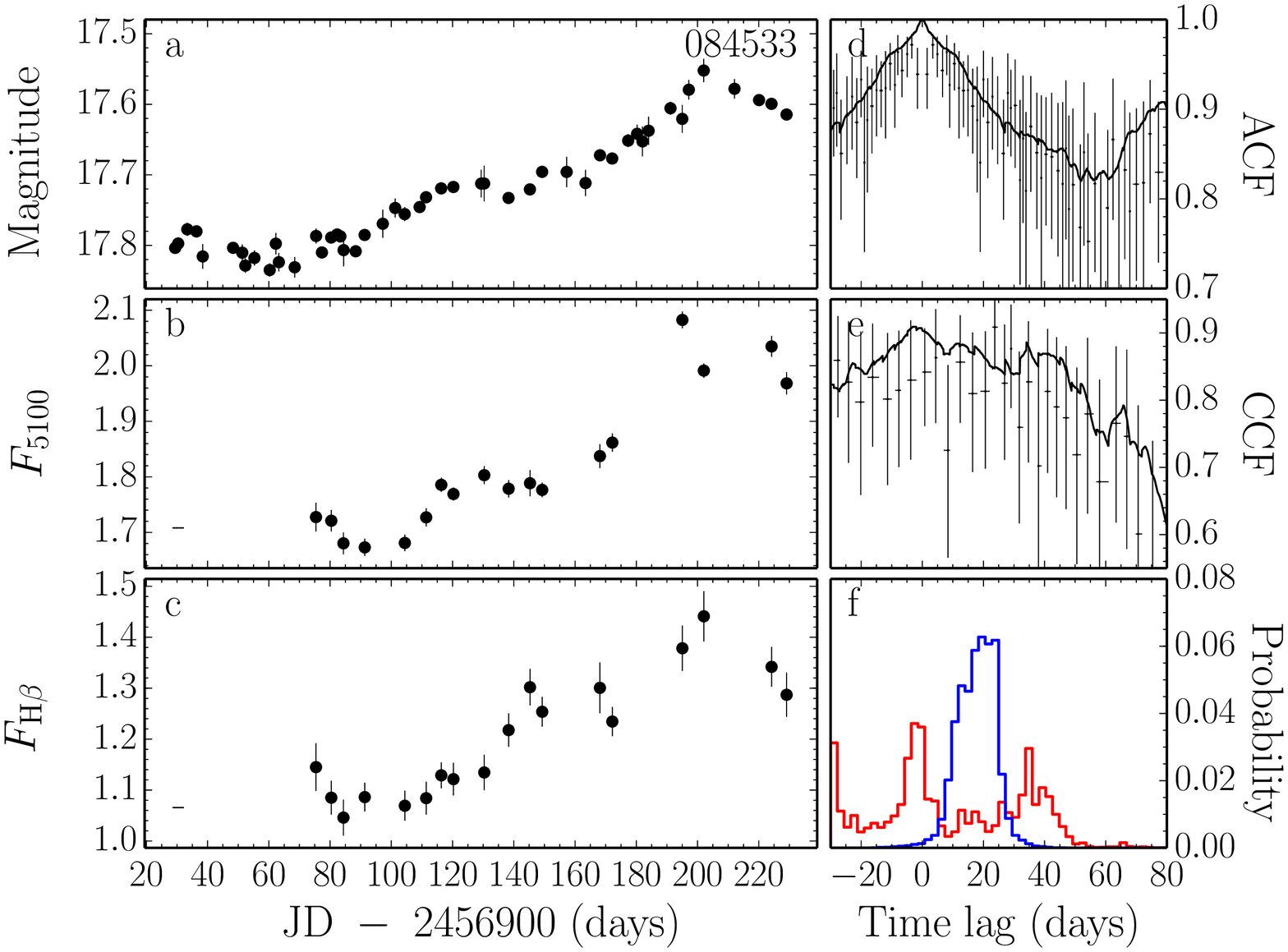}\hspace{0.5cm}
\includegraphics[angle=0,width=0.48\textwidth]{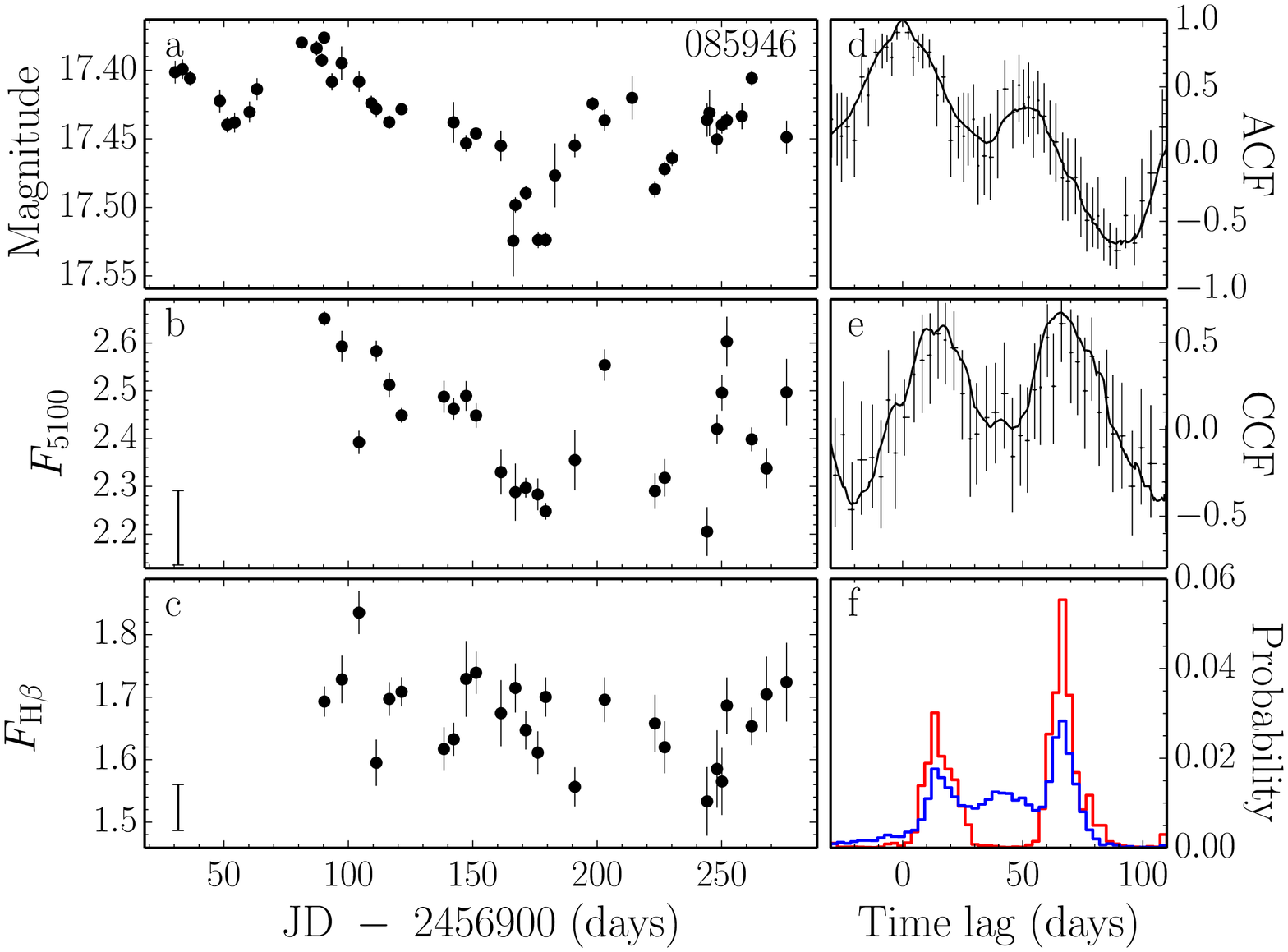}
\end{center}
\vspace{-0.5cm}
\caption{\footnotesize Light curves and cross-correlation results. Each object has six panels: 
    ({\it a, b, c}) are light curves of SDSS $r^{\prime}$-band magnitude, 5100 \AA\ continuum and H$\beta$ 
emission, respectively; ({\it d, e, f}) are auto-correlation function 
(ACF) of the $r^{\prime}$-band magnitude (5100 \AA\ continuum for J080131; 
see the main text), cross-correlation function (CCF) of the $r^{\prime}$-band 
magnitude and H$\beta$ line emission (5100 \AA\ and H$\beta$ for J080131)
and the Monte Carlo simulations of peak (red) and 
centroid (blue) of the lags, respectively. In panels {\it d} and {\it e}, the solid lines 
show the results of the ICCF method, and the points with error bars are from ZDCF ($Z-$transformed 
discrete correlation function). $F_{5100}$ and $F_{\rm H\beta}$ are in units of 
$10^{-16}{\rm erg~s^{-1}~cm^{-2} \AA^{-1}}$ and $10^{-14}{\rm erg~s^{-1}~cm^{-2}}$, 
respectively. Bars with terminals denote systematic errors and are plotted in the corners of the panels
(see Paper I for details). For J084533, the systematic error bars are so small that the caps of error 
bars merge in panels {\it b} and {\it c}; the same holds for J102339 in panel {\it b}. 
}
\end{figure*}

\subsection{Host Galaxies}
Like the SEAMBH2013 sample, we have no observations that can clearly separate the host galaxies 
of the AGNs in SEAMBH2014. 
Shen et al. (2011) propose the following empirical 
relation to estimate the fractional contribution of the host galaxy to the optical continuum 
emission: $L_{5100}^{\rm host}/L_{5100}^{\rm AGN}=0.8052-1.5502x+0.912x^2-0.1577x^3$, for $x<1.053$, 
where $x=\log \left(L_{5100}^{\rm tot}/10^{44}{\rm erg~s^{-1}}\right)$ and 
$L_{5100}^{\rm tot}$ is the total emission from the AGN and its host at 5100 \AA. For  $x>1.053$, 
$L_{5100}^{\rm host}\ll L_{5100}^{\rm AGN}$, and the host contamination can be neglected.
The host fractions at 5100 \AA\ for the objects (J075949, J080131, J084533, J085946 and J102339) 
are (27.5\%, 37.1\%, 14.2\%, 19.0\% and 31.8\%). The values of $L_{5100}$ listed in Table 5 are 
the host-subtracted luminosities.
We note that this empirical relation is based on SDSS spectroscopic observations with a 
$3^{\prime\prime}$ fiber, whereas we used a $5^{\prime\prime}$-wide slit. It should apply to our
observations reasonably well (see Paper IV for additional discussions on this issue). 
We will revisit this issue in the future using high-resolution images that can more reliably separate the host.

\section{Measurements of H$\beta$ Lags, Black Hole Masses and Accretion Rates}
\subsection{Lags}
As in Papers I--IV, we used cross-correlation analysis to determine H$\beta$ lags relative 
to photometric or 5100 \AA\, continuum light curves. We use the centroid lag for H$\beta$.
The uncertainties on the lags are determined 
through the ``flux randomization/random subset sampling" method (RS/RSS; Peterson et al. 1998, 
2004). The cross-correlation centroid distribution (CCCD, described in Appendix E) 
and cross-correlation peak distribution 
(CCPD) generated by the FR/RSS method (Maoz \& Netzer 1989; Peterson et al. 1998, 2004; Denney et al. 
2006, 2010, and references therein) are shown in Figure 1. We used the following criteria to define 
a successful detection of H$\beta$ lag: 1) non-zero lag from the CCF peak and 2) a maximum 
correlation coefficient larger than 0.5.  Data for the light curves of the targets are given 
in Tables 2--4. All the measurements of the SEAMBH2014 sample are provided in Table 5.

The $r^{\prime}$-band light curves are generally consistent with the 5100 \AA\, continuum light 
curve, but the former usually have small scatter, as shown in Figure 1. We calculated CCFs for 
the H$\beta$ light curves with both $r^{\prime}$-band photometry and with 5100 \AA\, spectral 
continuum for all objects. The quality of the H$\beta-r^{\prime}$ CCFs is usually better 
than the H$\beta-F_{5100}$ CCFs. We show the H$\beta-r^{\prime}$ CCFs for all objects in Figure 1, 
except for J080131.  We use the H$\beta-r^{\prime}$ lags in the following analysis. 
For J080131, the $r^{\prime}$-band light curve between 200 and 220 days does not match the 5100 
\AA\, continuum light curve, even though H$\beta$ does follow 5100 \AA\, continuum tightly. Notes 
to individual sources are given in Appendix D.


\subsection{Black Hole Masses and Accretion Rates}
There are two ways of calculating BH mass, base either on the RMS spectrum (e.g., Peterson et al. 
2004; Bentz et al. 2009; Denney et al. 2010; Grier et al. 2012) or on the mean spectrum 
(e.g., Kaspi et al. 2005; Papers I--IV).  Different studies also adopt 
different measures of the line width, typically either the line dispersion 
$\sigma_{\rm line}$ (second moment of the line profile) or the FWHM.  In this 
study, we choose to parameterize the line width using FWHM, as measured in the 
mean spectra. The narrow H$\beta$ component may influence the measurement of FWHM. 
We adopt the same procedure as in Paper I
to remove the narrow H$\beta$. We fix narrow H$\beta$/\oiii$\lambda5007$ to 0.1, and measure 
FWHM from the mean spectra with narrow H$\beta$ subtracted. Then we set 
 H$\beta$/\oiii$\lambda5007$ to 0 and 0.2 and repeat the process to obtain 
lower and upper limits to FWHM.
The relatively wide slit employed in our campaign ($5^{\prime\prime}$) 
significantly broadens the emission lines by $V_{\rm inst} \approx 1200\,\kms$, 
where $V_{\rm inst}$ is the instrumental broadening that can be estimated from the broadening 
of selected comparison stars. As in Paper IV, we obtain the intrinsic 
width of the mean spectra from ${\rm FWHM}=\left({\rm FWHM}_{\rm obs}^2-V_{\rm inst}^2\right)^{1/2}$. 
The FWHM simply obtained here is accurate enough for BH mass estimation. 
Our procedure for BH mass estimation is based on FWHM measured from the mean spectrum (see explanation 
in Papers I and II). As shown recently in Woo et al. (2015), the scatter in the scaling parameter 
($f_{\rm BLR}$) derived in this method is very similar to the scatter in the method based on the RMS 
spectrum.  We use Equations (2) and (3) to calculate accretion rates and BH masses for the five 
sources listed in Table 5. For convenience and completeness, Table 5 also lists H$\beta$ lags, 
BH masses and accretion rates for the sources from SEAMBH2012 and SEAMBH2013. Our campaign has successfully 
detected H$\beta$ lags for 18 SEAMBHs since October 2012.

\begin{figure}
\begin{center}
\includegraphics[angle=0,width=0.48\textwidth]{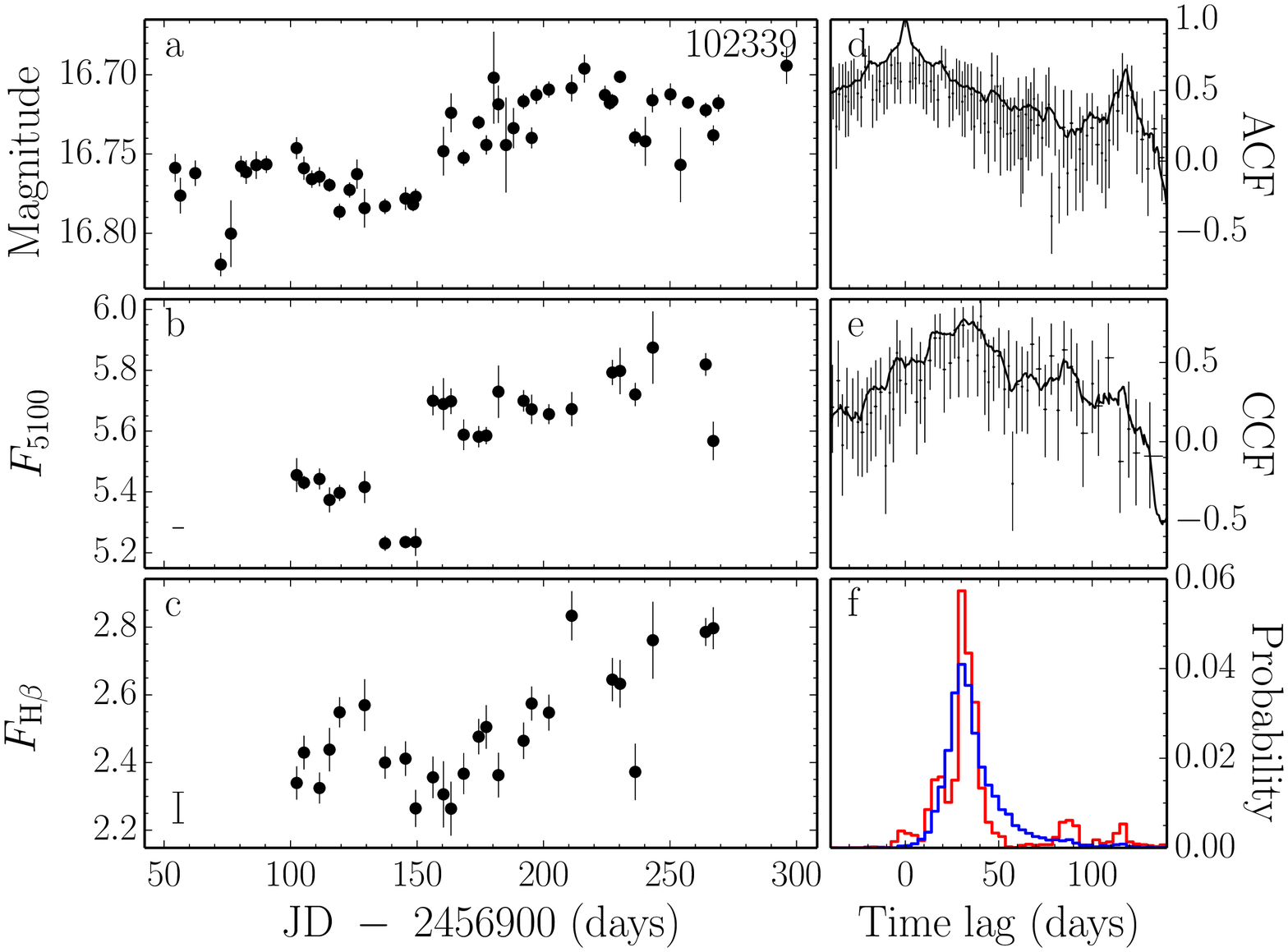}
\end{center}
{Figure 1 {\it continued.}}
\end{figure}

As described in Paper II, there are some theoretical uncertainties in identifying a critical value of 
$\mathdotM$ to define a SEAMBH (Laor \& Netzer 1989; Beloborodov 1998; Sadowski et al. 2011). Following 
Paper II, we classified SEAMBHs as those objects with $\eta\mathdotM \geq 0.1$.
This is based on the idea that beyond this value, the accretion disk becomes 
slim and the radiation efficiency is reduced mainly due to photon trapping (Sadowski et al. 2011). 
Since we currently cannot observe the entire spectral energy distribution, we have no direct way to 
measure $\er$, and this criterion is used as an approximate tool 
to identify SEAMBH candidates. To be on the conservative side, we chose the lowest possible 
efficiency, $\eta=0.038$ (retrograde disk with $a=-1$; see Bardeen et al. 1972). Thus, SEAMBHs 
are objects with $\mathdotM=2.63$. For simplicity, in this paper we use $\mathdotM_{\rm min}=3$ 
as the required minimum (Papers II and IV). We refer to AGNs with $\mathdotM\ge 3$ as SEAMBHs and those with 
$\mathdotM<3$ as sub-Eddington ones.  Paper IV clearly shows that the properties of the $\rblrl$ relation
for $\mathdotM\ge 3$ and $\mathdotM<3$ are significantly different.

\begin{deluxetable}{rlcrll}[!b]
  \tablecolumns{6}
  \setlength{\tabcolsep}{3pt}
  \tablewidth{0pc}
  \tablecaption{Light curves of J102339}
  \tabletypesize{\scriptsize}
  \tablehead{
      \multicolumn{6}{c}{J102339}        \\ \cline{1-6}
      \multicolumn{2}{c}{Photometry}     &
      \colhead{}                         &
      \multicolumn{3}{c}{Spectra}        \\ \cline{1-2}\cline{4-6}
      \colhead{JD}                       &
      \colhead{mag}                      &
      \colhead{}                         &
      \colhead{JD}                       &
      \colhead{$F_{5100}$}               &
      \colhead{$F_{\rm H\beta}$}                  }
\startdata
  54.400 & $16.759\pm 0.009$  & &  102.431 & $ 5.455\pm 0.056$ & $ 2.340\pm 0.049$ \\
  56.431 & $16.776\pm 0.011$  & &  105.309 & $ 5.430\pm 0.022$ & $ 2.429\pm 0.050$ \\
  62.333 & $16.762\pm 0.008$  & &  111.450 & $ 5.443\pm 0.034$ & $ 2.325\pm 0.046$ \\
  72.412 & $16.820\pm 0.007$  & &  115.406 & $ 5.374\pm 0.041$ & $ 2.438\pm 0.064$ \\
  76.438 & $16.800\pm 0.021$  & &  119.396 & $ 5.397\pm 0.027$ & $ 2.548\pm 0.044$ \\
  \enddata
  \tablecomments{\footnotesize
      This table is available in its entirety in a machine-readable form in the online journal. A portion is shown here for guidance regarding its form and content.
      }
\end{deluxetable}

Figure 2 plots distributions of $L_{5100}$, EW(H$\beta$), $\mathdotM$ and $\bhm$ of all the mapped AGNs 
(41 from Bentz et al. 2013 and the 18 SEAMBHs from our campaign; see Table 7 in Paper IV and Table 5 here).  
As shown clearly in the diagrams, SEAMBH targets are generally more luminous by a factor of 2--3 compared 
to previous RM AGNs (Figure 2{\it a}).  The BH masses of SEAMBHs are generally less smaller by a factor of 
10 compared to previous samples, whereas, as a consequence of our selection, the accretion rates of SEAMBHs 
are higher by 2--3 orders of magnitude (Figures 2{\it c} and 2{\it d}).  However, EW(H$\beta$) of SEAMBHs 
are not significantly smaller (Figure 2{\it d}). On average, the high$-\mathdotM$ sources have lower mean 
EW(H$\beta$), consistent with the inverse correlation between EW(H$\beta$) and $L_{\rm bol}/L_{\rm Edd}$ 
(e.g., Netzer et al. 2004).

\renewcommand{\arraystretch}{1.5}
\begin{deluxetable*}{lccccccrccc}
 \tablecolumns{9}
 \setlength{\tabcolsep}{3pt}
\tablewidth{\textwidth}
 \tablecaption{H$\beta$ Reverberations of the SEAMBHs}
 \tabletypesize{\scriptsize}
 \tablehead{
     \colhead{Objects}                  &
     \colhead{$\tauhb$}                 &
     \colhead{FWHM}                     &
     \colhead{$\sigma_{\rm line}$}      &
     \colhead{$\log \left(\bhm/\sunm\right)$}&
     \colhead{$\log \mathdotM$}         &
     \colhead{$\log L_{5100}$}          &
     \colhead{$\log \Lhb$}              &
     \colhead{EW(H$\beta$)}             \\ \cline{2-9}
     \colhead{}                         &
     \colhead{(days)}                   &
     \colhead{($\kms$)}                 &
     \colhead{($\kms$)}                 &
     \colhead{}                         &
     \colhead{}                         &
     \colhead{($\ergs)$}                &
     \colhead{($\ergs)$}                &
     \colhead{(\AA)}
 }
\startdata
 \multicolumn{9}{c}{SEAMBH2012}\\ \hline
          Mrk 335  & $  8.7_{- 1.9}^{+ 1.6} $ & $ 2096\pm170 $ & $1470\pm 50$ & $ 6.87_{-0.14}^{+0.10} $ & $  1.28_{-0.30}^{+0.37} $ & $ 43.69\pm 0.06 $ & $ 42.03\pm 0.06 $ & $ 110.5\pm 22.3 $ \\ 
         Mrk 1044  & $ 10.5_{- 2.7}^{+ 3.3} $ & $ 1178\pm 22 $ & $ 766\pm  8$ & $ 6.45_{-0.13}^{+0.12} $ & $  1.22_{-0.41}^{+0.40} $ & $ 43.10\pm 0.10 $ & $ 41.39\pm 0.09 $ & $ 101.4\pm 31.9 $ \\ 
          Mrk 382  & $  7.5_{- 2.0}^{+ 2.9} $ & $ 1462\pm296 $ & $ 840\pm 37$ & $ 6.50_{-0.29}^{+0.19} $ & $  1.18_{-0.53}^{+0.69} $ & $ 43.12\pm 0.08 $ & $ 41.01\pm 0.05 $ & $  39.6\pm  9.0 $ \\ 
          Mrk 142  & $  7.9_{- 1.1}^{+ 1.2} $ & $ 1588\pm 58 $ & $ 948\pm 12$ & $ 6.59_{-0.07}^{+0.07} $ & $  1.65_{-0.23}^{+0.23} $ & $ 43.56\pm 0.06 $ & $ 41.60\pm 0.04 $ & $  55.2\pm  9.5 $ \\ 
      IRAS F12397  & $  9.7_{- 1.8}^{+ 5.5} $ & $ 1802\pm560 $ & $1150\pm122$ & $ 6.79_{-0.45}^{+0.27} $ & $  2.26_{-0.62}^{+0.98} $ & $ 44.23\pm 0.05 $ & $ 42.26\pm 0.04 $ & $  54.2\pm  8.4 $ \\ 
          Mrk 486  & $ 23.7_{- 2.7}^{+ 7.5} $ & $ 1942\pm 67 $ & $1296\pm 23$ & $ 7.24_{-0.06}^{+0.12} $ & $  0.55_{-0.32}^{+0.20} $ & $ 43.69\pm 0.05 $ & $ 42.12\pm 0.04 $ & $ 135.9\pm 20.3 $ \\ 
          Mrk 493  & $ 11.6_{- 2.6}^{+ 1.2} $ & $  778\pm 12 $ & $ 513\pm  5$ & $ 6.14_{-0.11}^{+0.04} $ & $  1.88_{-0.21}^{+0.33} $ & $ 43.11\pm 0.08 $ & $ 41.35\pm 0.05 $ & $  87.4\pm 18.1 $ \\ 
       IRAS 04416  & $ 13.3_{- 1.4}^{+13.9} $ & $ 1522\pm 44 $ & $1056\pm 29$ & $ 6.78_{-0.06}^{+0.31} $ & $  2.63_{-0.67}^{+0.16} $ & $ 44.47\pm 0.03 $ & $ 42.51\pm 0.02 $ & $  55.8\pm  4.7 $ \\  \hline
 \multicolumn{9}{c}{SEAMBH2013}\\ \hline
     SDSS J075101  & $ 33.4_{- 5.6}^{+15.6} $ & $ 1495\pm 67 $ & $1055\pm 32$ & $ 7.16_{-0.09}^{+0.17} $ & $  1.34_{-0.41}^{+0.25} $ & $ 44.12\pm 0.05 $ & $ 42.25\pm 0.03 $ & $  68.1\pm  8.6 $ \\ 
     SDSS J080101  & $  8.3_{- 2.7}^{+ 9.7} $ & $ 1930\pm 18 $ & $1119\pm  3$ & $ 6.78_{-0.17}^{+0.34} $ & $  2.33_{-0.72}^{+0.39} $ & $ 44.27\pm 0.03 $ & $ 42.58\pm 0.02 $ & $ 105.5\pm  8.3 $ \\ 
     SDSS J080131  & $ 11.5_{- 3.6}^{+ 8.4} $ & $ 1188\pm  3 $ & $ 850\pm 12$ & $ 6.50_{-0.16}^{+0.24} $ & $  2.46_{-0.54}^{+0.38} $ & $ 43.98\pm 0.04 $ & $ 42.08\pm 0.03 $ & $  64.0\pm  7.0 $ \\ 
     SDSS J081441  & $ 18.4_{- 8.4}^{+12.7} $ & $ 1615\pm 22 $ & $1122\pm 11$ & $ 6.97_{-0.27}^{+0.23} $ & $  1.56_{-0.57}^{+0.63} $ & $ 44.01\pm 0.07 $ & $ 42.42\pm 0.03 $ & $ 132.0\pm 23.7 $ \\ 
     SDSS J081456  & $ 24.3_{-16.4}^{+ 7.7} $ & $ 2409\pm 61 $ & $1438\pm 32$ & $ 7.44_{-0.49}^{+0.12} $ & $  0.59_{-0.30}^{+1.03} $ & $ 43.99\pm 0.04 $ & $ 42.15\pm 0.03 $ & $  74.4\pm  7.6 $ \\ 
     SDSS J093922  & $ 11.9_{- 6.3}^{+ 2.1} $ & $ 1209\pm 16 $ & $ 835\pm 11$ & $ 6.53_{-0.33}^{+0.07} $ & $  2.54_{-0.20}^{+0.71} $ & $ 44.07\pm 0.04 $ & $ 42.09\pm 0.04 $ & $  53.0\pm  6.7 $ \\ \hline
 \multicolumn{9}{c}{SEAMBH2014}\\ \hline
     SDSS J075949  & $ 55.0_{-13.1}^{+17.0} $ & $ 1807\pm 11 $ & $1100\pm  3$ & $ 7.54_{-0.12}^{+0.12} $ & $  0.70_{-0.29}^{+0.29} $ & $ 44.20\pm 0.03 $ & $ 42.48\pm 0.02 $ & $  97.5\pm  9.1 $ \\ 
     SDSS J080131  & $ 11.2_{- 9.8}^{+14.8} $ & $ 1290\pm 13 $ & $ 800\pm  5$ & $ 6.56_{-0.90}^{+0.37} $ & $  2.29_{-0.80}^{+1.87} $ & $ 43.95\pm 0.04 $ & $ 41.96\pm 0.05 $ & $  52.3\pm  7.7 $ \\ 
     SDSS J084533  & $ 15.2_{- 6.3}^{+ 3.2} $ & $ 1243\pm 13 $ & $ 818\pm 10$ & $ 6.66_{-0.23}^{+0.08} $ & $  2.98_{-0.22}^{+0.52} $ & $ 44.54\pm 0.04 $ & $ 42.58\pm 0.05 $ & $  55.9\pm  7.5 $ \\ 
 SDSS J085946  & $ 34.8_{-26.3}^{+19.2} $ & $ 1718\pm 16 $ & $1031\pm 14$ & $ 7.30_{-0.61}^{+0.19} $ & $  1.51_{-0.43}^{+1.27} $ & $ 44.41\pm 0.03 $ & $ 42.51\pm 0.02 $ & $  63.1\pm  5.2 $ \\ 
     SDSS J102339  & $ 24.9_{- 3.9}^{+19.8} $ & $ 1733\pm 29 $ & $1139\pm 19$ & $ 7.16_{-0.08}^{+0.25} $ & $  1.29_{-0.56}^{+0.20} $ & $ 44.09\pm 0.03 $ & $ 42.14\pm 0.03 $ & $  57.0\pm  5.9 $ 
\enddata
\tablecomments{\footnotesize 
All SEAMBH2012 measurements are taken from Paper III, but 5100 \AA\, fluxes are from 
I and II, SEAMBH2013 from Paper IV, and SEAMBH2014 is the present paper.
MCG +06$-$26$-$012 was selected as a super-Eddington candidate in SEAMBH2012 but later was identified 
to be a sub-Eddington accretor ($\mathdotM=0.46$); we discard it here.
}
\end{deluxetable*}

\section{Properties of H$\beta$ Lags in SEAMBHs}
The $\rhb-L_{5100}$ correlation was originally presented by Peterson (1993; his Figure 10, only nine objects). 
It was confirmed by Kaspi et al. (2000) using a sample of 17 low-redshift quasars. Bentz et al. (2013) refined 
the $\rblrl$ relation through subtraction of host contamination and found that its intrinsic scatter is only 
0.13 dex. Paper IV (Table 7) provides a complete list of previously mapped AGNs, based on Bentz et al. (2013); 
we directly use these values\footnote{NGC 7469 was mapped twice by Collier et al. (1998) and Peterson et al. 
(2014). While their H$\beta$ lags are consistent, the FWHM of H$\beta$ is very different. We only retain the 
later observation in the analysis.}. As in Paper IV, for
objects with multiple measurements of H$\beta$ lags, we obtain the BH mass from each campaign and then calculate 
the average BH mass. Using the averaged BH mass, we apply it to get accretion rates of the BHs during each 
monitoring epoch, which are further averaged to obtain the mean accretion rates of those objects (Kaspi et al. 
2005; Bentz et al. 2013).  We call this the ``average scheme."  On the other hand,
we may consider each individual measurement of a single object as different objects (e.g., Bentz et 
al. 2013). We called this the ``direct scheme." Although the two approaches are in principle different,
we obtain very similar results (see a comparison in Paper IV).  

All correlations of two parameters shown in this paper are calculated with the FITEXY method, using the 
version adopted by Tremaine et al. (2002), which allows for intrinsic scatter by 
increasing the uncertainties in small steps until $\chi^2$ reaches unity (this is typical for 
many of our correlations). We also emplot the BCES method (Akristas \& Bershady 1996) but prefer 
not to use its results because it is known to give unreliable results in samples containing 
outliers (there are a few objects with quite large uncertainties of $\mathdotM$).

\subsection{The $\rblrl$ relation}
As shown in Paper IV, the H$\beta$ lags of the SEAMBH2013 sample were found to significantly 
deviate from the normal $\rblrl$ relation, by a factor of a few. We plot the $\rblrl$ relation 
of all samples in Figure 3. For sub-Eddington AGNs ($\mathdotM\le 3$) in the direct scheme, 
$\log \left(\rhb/{\rm ltd}\right)=(1.54\pm0.03)+(0.53\pm0.03)\log \ell_{44},$
with an intrinsic scatter of $0.15$ (see Paper IV). Using FITEXY, we have
\begin{equation}
\begin{split}
&\hspace{-0.5cm}\log \left(\rhb/{\rm ltd}\right)=\\
&\left\{\begin{array}{ll}
(1.30\pm 0.05)+(0.53\pm 0.06)\log \ell_{44} & (\mathdotM\ge 3),\\ [0.8em]
(1.44\pm 0.03)+(0.49\pm 0.03)\log \ell_{44} & ({\rm for~ all~\mathdotM}),
\end{array}\right.
\end{split}
\end{equation}

\noindent with intrinsic scatters of $\sigma_{\rm in}=(0.24,0.21)$. Clearly, the intrinsic scatter 
of SEAMBHs is much larger than the sample of sub-Eddington AGNs. In the averaged scheme, 
we have
$\log \left(\rhb/{\rm ltd}\right)=(1.55\pm0.04)+(0.53\pm0.04)\log \ell_{44}$\,
for sub-Eddington AGNs, with an intrinsic scatter of $0.16$ 
(see Paper IV), and 

%
\begin{equation}
\begin{split}
&\hspace{-0.5cm}\log \left(\rhb/{\rm ltd}\right)= \\
&\left\{\begin{array}{ll}
(1.32\pm 0.05)+(0.52\pm 0.06)\log \ell_{44} & (\mathdotM\ge 3),\\ [0.8em]
(1.44\pm 0.03)+(0.49\pm 0.03)\log \ell_{44} & ({\rm for~ all~\mathdotM}),
\end{array}\right.
\end{split}
\end{equation}

\noindent with intrinsic scatters of $\sigma_{\rm in}=(0.22,0.21)$.
The slope of the correlation for the SEAMBH sample is comparable 
to that of sub-Eddington AGNs, but the normalization is significantly different. It is clear 
that the SEAMBH sources increase the scatter considerably, especially over the limited 
luminosity range occupied by the new sources. 

As in Paper IV, we also tested the correlation between H$\beta$ lag and H$\beta$ luminosity, 
namely, the $\rhb-\Lhb$ relation. The scatter of the $\rhb-\Lhb$ correlation is not smaller 
than that of the $\rhb-L_{5100}$ correlation, and we do not consider it further.

\begin{figure*}[t!]
\begin{center}
\includegraphics[angle=0,width=1.0\textwidth]{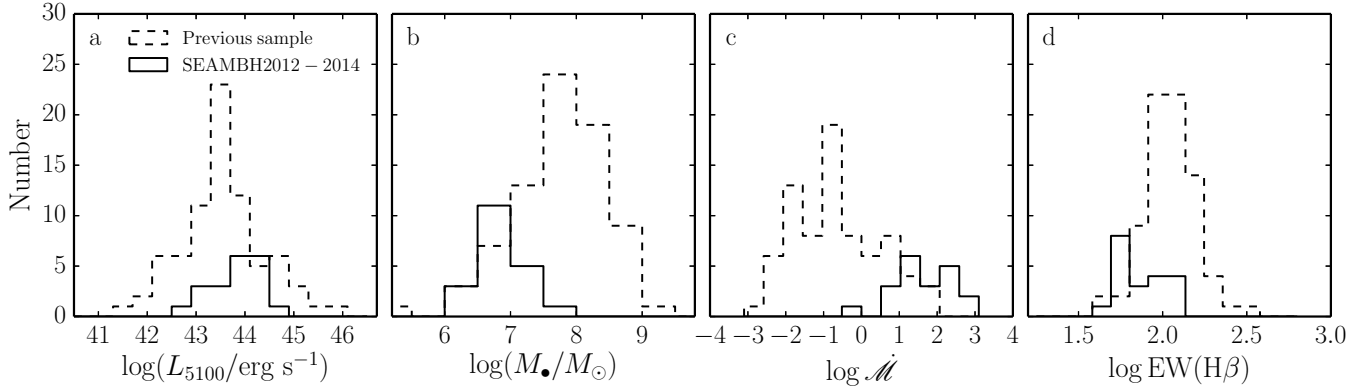}
\end{center}
\vspace{-0.5cm}
\caption{\footnotesize 
Distributions of 5100 \AA\, luminosity ($L_{5100}$), BH mass ($\bhm$), dimensionless 
accretion rate ($\mathdotM$), and equivalent width (EW) of all the mapped AGNs. 
These distributions show that the present sample of mapped AGNs is inhomogeneous.
Only three luminous sources ($L_{5100}\gtrsim 10^{45}\ergs$) have been mapped.
The distribution of EW(H$\beta$) in panel {\it d}\ shows that the SEAMBH sample tends to 
have low EW(H$\beta$).
}
\label{dotm}
\end{figure*}

\subsection{$\mathdotM-$dependent BLR Size}

To test the dependence of the BLR size on accretion rate, we define a new parameter,  
$\Delta \rhb=\log\left(\rhb/\prblr\right)$,
that specifies the deviation of individual objects from the $\rblrl$ relation of the 
subsample of $\mathdotM<3.0$ sources (i.e., $\prblr$ as given by Equations 4b and 5b for
$\mathdotM<3$ AGNs in Paper IV). The scatter of $\Delta \rhb$ is calculated by  
$\sigma_{_{\rhb}}=\left[\sum_i\left(\Delta R_{{\rm H\beta}, i}-\langle\Delta \rhb\rangle\right)^2/N\right]^{1/2}$, 
where $N$ is the number of objects and $\langle \Delta R_{{\rm H\beta}}\rangle$ is the averaged 
value. Figure 3 provides $\Delta\rhb$ plots for comparison.

Figure 4 shows $\Delta \rhb$ versus $\mathdotM$, as well as $\Delta \rhb$ distributions
for the $\mathdotM\ge3$ and $\mathdotM<3$ subsamples in the direct (panels a and b) and averaged
(panels c and d) schemes. A Kolmogorov-Smirnov (KS) test 
shows that the probability that the two subsamples are drawn from the same parent distributions 
is $p_{_{\rm KS}}=0.00029$ for the direct scheme and $p_{_{\rm KS}}=0.0094$ for the averaged
scheme. This provides a strong 
indication that the main cause of deviation from the normal $\rblrl$ relation
is the extreme accretion rate. Thus, a single $\rblrl$
relation for all AGNs is a poor approximation for a more complex situation in which both the 
luminosity and the accretion rate determine $R_{{\rm H\beta}}$. From the regression for 
$\mathdotM\ge 3$ AGNs, 
we obtain the dependence of the deviations of $\rhb$ from the $\rblrl$ relation in Figure 4:
%
\begin{equation}
\begin{split}
&\Delta \rhb=\\
&\left\{\begin{array}{ll}
(0.39\pm0.09)-(0.47\pm0.06)\log \mathdotM & ({\rm direct~scheme}),\\ [0.8em]
(0.34\pm0.09)-(0.42\pm0.07)\log \mathdotM & ({\rm averaged ~scheme}),
\end{array}\right.
\end{split}
\end{equation}
\vspace{0.3cm}

\noindent with $\sigma_{\rm in}=(0.01,0.05)$, respectively. 
We have tested the above correlations also for $\mathdotM<3$. 
The FITEXY regressions give slopes near 0, with very large uncertainties: 
$\Delta \rhb\propto \mathdotM^{-0.055\pm 0.032}$ and 
$\Delta \rhb\propto \mathdotM^{-0.095\pm0.050}$ for Figure 4a and 4c, respectively, implying 
that $\Delta\rhb$ does not correlate with $\mathdotM$ for the $\mathdotM<3$ group. All this confirms 
that $\mathdotM$ is an additional parameter that controls the $\rblrl$ relation in AGNs with 
high accretion rates.

\begin{figure*}[t!]
\begin{center}
\includegraphics[angle=0,width=0.80\textwidth]{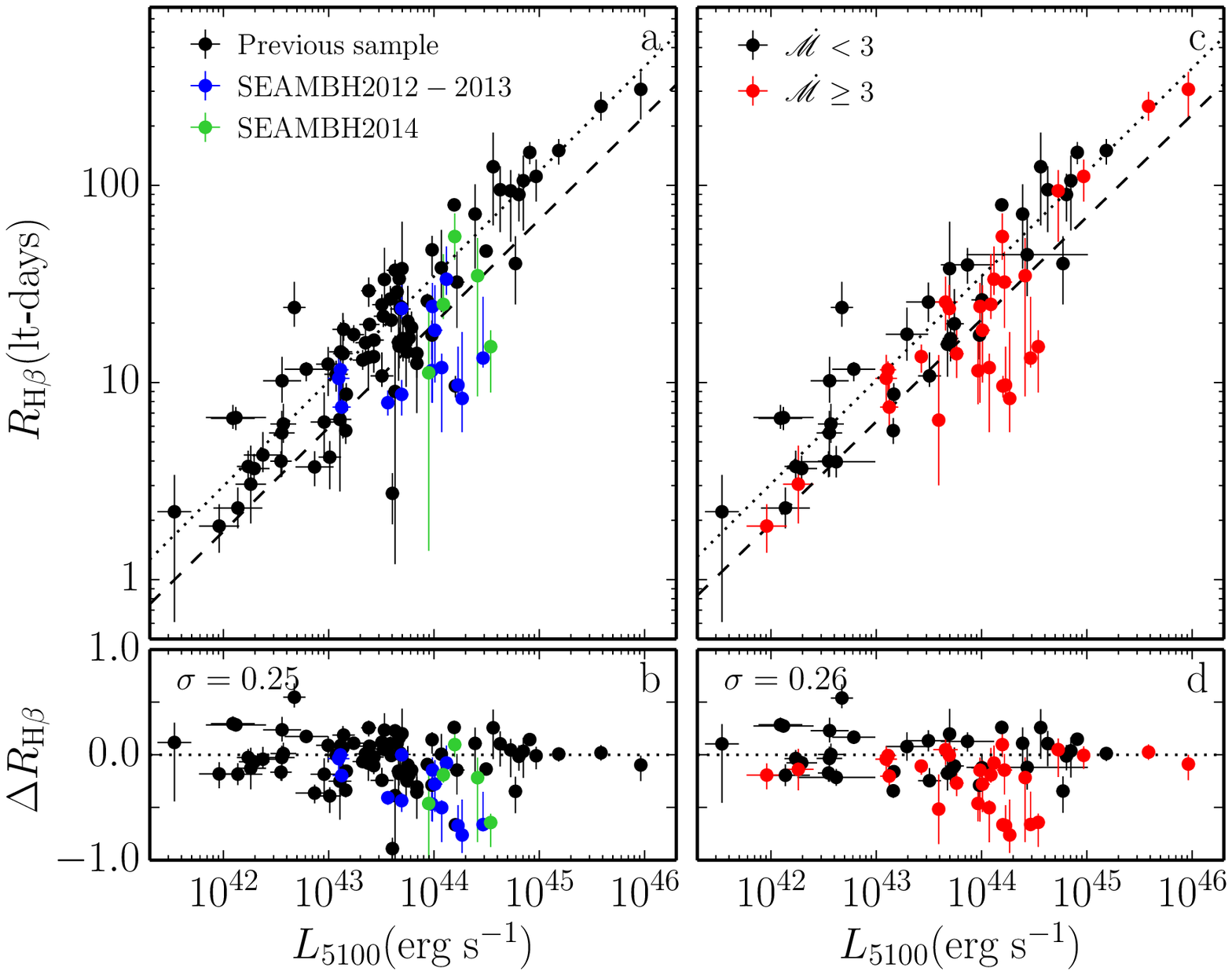}
\end{center}
\vspace{-0.5cm}
\caption{\footnotesize 
The $\rblrl$ plot for all mapped AGNs. {\it Left} panel shows multiple-RM results as individual 
points, whereas the {\it right} panel shows the averaged results of AGNs with multiple-RM measurements. 
The dotted line is the regression of $\rblrl$ relation for $\mathdotM<3$ AGNs (Equation 4); the
dashed line is the regression for the $\mathdotM\ge3$ objects. The scatter (standard deviations) of 
$\Delta R_{\rm H\beta}$ is given in the upper left corner of Panels {\it b} and {\it d}. 
}
\label{dotm}
\end{figure*}

\section{A New Scaling Relation for the BLR}
We provide evidence H$\beta$ lags depend on luminosity {\it and}\ accretion 
rate. 
There are a total of 
28 SEAMBHs (including those discovered in other studies).  We now have an 
opportunity to define a new scaling relation for the BLR, one that 
properly captures the behavior of sub-Eddington and super-Eddington AGNs.
Considering the dependence of $\Delta \rhb\propto \mathdotM^{-0.42}$ (Equation 6),
a unified form of the new scaling law can take the form\footnote{We have tried 
$\rhb=\alpha_1 \ell_{44}^{\beta_1}\left[1+\left(\mathdotM/\mathdotM_c\right)^{\gamma_1}\right]^{\delta_1}$, 
which is continuous for the transition from sub- to super-Eddington sources.
The fitting also yields a very rapid transition at $\mathdotM_{\rm c}\sim10$, 
with $\gamma_1 = 0.025$ and $\delta_1 = 21.02$ (the present sample is still dominated by sub-Eddington 
AGNs, with a ratio of 35/63). We prefer the form given by Equation (7). 
}
\begin{equation}
\rhb=\alpha_1\, \ell_{44}^{\beta_1}
     \min\left[1,\left(\frac{\mathdotM}{\mathdotM_{\rm c}}\right)^{-\gamma_1}\right],
\end{equation}
where $\mathdotM_c$ is to be determined by data. Equation (7) reduces to the normal $\rblrl$ relation 
for sub-Eddington AGNs and to 
$\rhb=\alpha_1\ell_{44}^{\beta_1}\left(\mathdotM/\mathdotM_c\right)^{-\gamma_1}$
for AGNs with $\mathdotM\ge \mathdotM_c$. There are four parameters to describe the new scaling relation, 
but only two ($\mathdotM_c$ and $\gamma_1$) are new due to the inclusion of accretion rates; the other 
two are mainly determined by sub-Eddington AGNs. The critical value of $\mathdotM_c$, which is different 
from the criterion of SEAMBHs, depends on the sample of SEAMBHs.

\begin{figure*}
\begin{center}
\includegraphics[angle=0,width=0.9\textwidth]{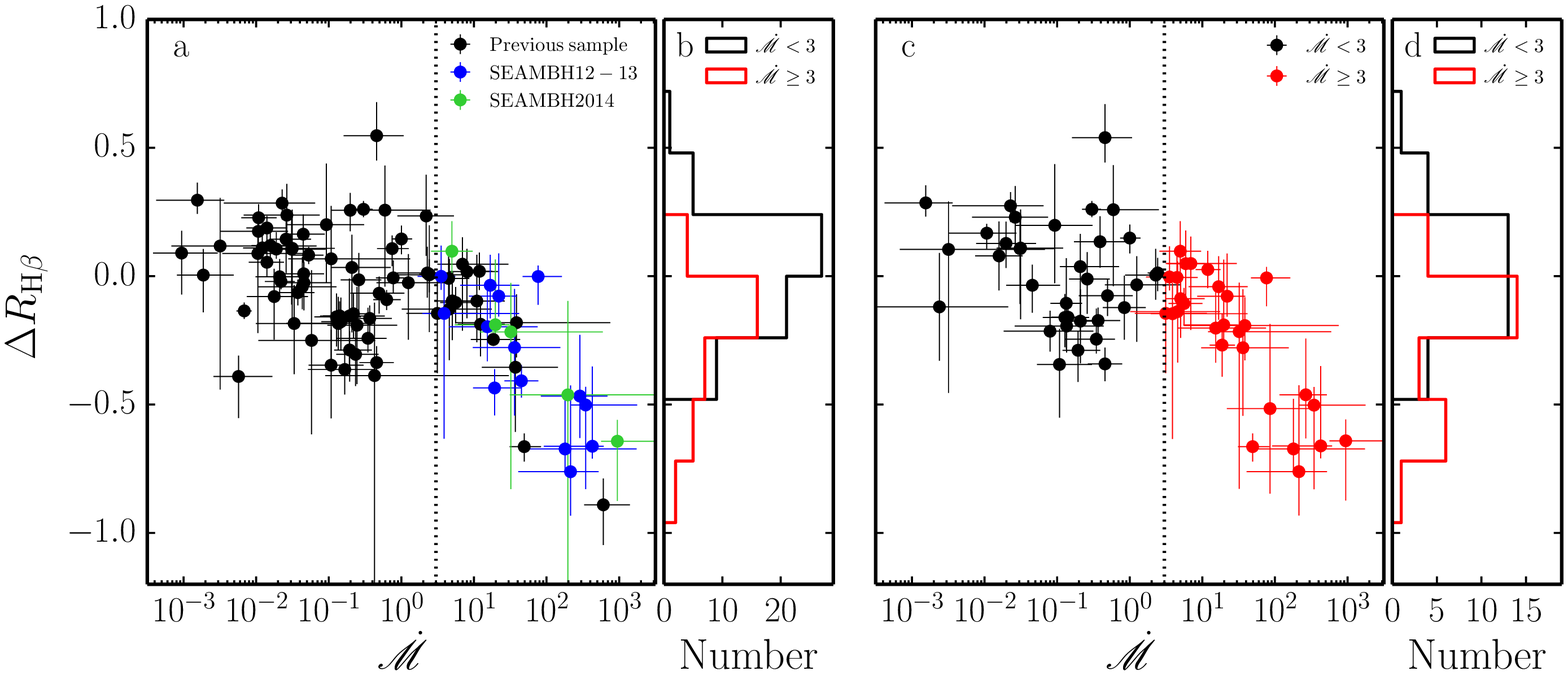}
\end{center}
\vspace{-0.5cm}
\caption{\footnotesize 
The deviation of H$\beta$ lags from the normal $\rblrl$ relation for sub-Eddington AGNs.
Panel {\it a} shows the sample of all mapped AGNs (the repeatedly monitored AGNs are regarded 
as individual ones). Panel {\it b} gives the distribution of low- and high-accretion 
objects. Panels {\it c} and {\it d} are the same plots but for the average scheme. Sub-Eddington 
AGNs show a random distributions, but SEAMBHs correlate with accretion rates. 
We note 
that $\bhm$ and $\mathdotM$ are calculated in exactly the same way for all objects, as 
indicated in Table 7 in Paper IV (i.e. using Equations 2 and 3).
}
\label{dotm}
\end{figure*}

In order to determine the four parameters simultaneously, we define
\begin{equation}
\chi^2=\frac{1}{N}\sum_{i=1}^N\frac{\left(\rhb-\rhb^i\right)^2}{\left(\Delta_{\rhb}^i\right)^2},
\end{equation}
where $\Delta_{\rhb}^i$ is the error bar of $\rhb^i$. Minimizing $\chi^2$ among all the mapped AGNs 
and employing a bootstrap method, we have
\begin{equation}
\alpha_1=29.6_{-2.8}^{+2.7};~~~\beta_1=0.56_{-0.03}^{+0.03};~~~\gamma_1=0.52_{-0.16}^{+0.33};
~~~\mathdotM_c=11.19_{-6.22}^{+2.29}.
\end{equation}
This new empirical relation has a scatter of $0.19$, smaller than
the scatter (0.26) of the normal $\rblrl$ relation for all the mapped AGNs. 
The new scaling relation is plotted in Figure 5.

Equation (7) shows the dependence of the BLR size on accretion rates, but it 
cannot be directly applied to single-epoch spectra for BH mass without 
knowning $\mathdotM$. Iteration of Equation (7) does not converge.  The reason 
is due to the fact that larger $\mathdotM$ leads to smaller $\rhb$ and higher 
$\mathdotM$, implying that the iteration from Equation (1) does not converge. 
Du et al. (2016b) devised a new method to determine $\mathdotM$ from 
single-epoch spectra.  Beginning with the seminal work of Boroson \& Green 
(1992), it has been well-known that  $\RFe \equiv F_{\rm FeII}/F_{\rm H\beta}$,
the flux ratio of broad optical \feii\ to H$\beta$, correlates strongly with 
Eddington ratio (Sulentic et al. 2000; Shen \& Ho 2014).  At the same time, 
the shape of broad H$\beta$, as parameterized by $\Dhb={\rm FWHM}/\sighb$, 
where $\sighb$ is the line dispersion, also correlates with Eddington ratio 
(Collin et al. 2006).  Combining the two produces produces a strong bivariate 
correlation, which we call the fundamental plane of the BLR, of the form
\begin{equation}
\log\mathdotM=\alpha_2+\beta_2\Dhb+\gamma_2\RFe,
\end{equation}
where 
\begin{equation}
\alpha_2=2.47\pm0.34;~ \beta_2=-1.59\pm0.14;~ \gamma_2=1.34\pm0.20.
\end{equation}

\section{Discussion}

\subsection{Normalized BLR Sizes}
In order to explore the relation between BLR size and accretion rate,
we define a dimensionless radius for the BLR, $\rrhb=\rhb/\Rg$, where 
$\Rg=1.5\times 10^{12}m_7$\,cm is the gravitational radius.  As in Paper IV,  
we insert Equation (3) into $\rrhb$ to replace $\ell_{44}$, to obtain 
$\rrhb=1.9\times 10^4~\mathdotM^{0.35}m_7^{-0.29}$.
This relation implies that $\rrhb$ increases with accretion rates as 
$\rrhb\propto \mathdotM^{0.35}$ for sub-Eddington AGNs, whereas in SEAMBHs 
$\rrhb\propto \mathdotM^{0.29\pm0.08}$ (as shown in Figure 6{\it a})
and $\rrhb$ and tends toward a maximum 
saturated value of $\rrhb^{\rm max}=\fblr^{-1}\left(c/V_{\rm min}\right)^2
=9\times 10^4\,\fblr^{-1}V_{\rm min,3}^{-2}$, where $V_{\rm min,3}=V_{\rm min}/
10^3\kms$ is the minimum velocity width of H$\beta$ (see Equation 15 in Paper 
IV). We note that the minimum observed FWHM values of H$\alpha$ (which is 
comparable to H$\beta$) is $\sim 10^3\, \kms$ among low-mass AGNs (Greene \& 
Ho 2007; Ho \& Kim 2016).  Indeed, this limit is consistent with the saturation trend of 
$\rrhb$ (Figure 6{\it a}).

We note that the relatively large scatter in Figure 6{\it a} is mostly due to 
the uncertainties in BH mass. In order to better understand the relation 
between the BLR and the central engine, we define, as in Paper IV, the 
parameter $Y=m_7^{0.29}\rrhb$, which reduces to
\begin{equation}
Y=1.9\times 10^4~\mathdotM^{0.35}.
\end{equation}
We would like to point out that Equation (12) 
describes the coupled system of the BLR and the accretion disks. 
It is therefore expected that $Y$ is a synthetic parameter
describing the photoionization process including ionizing sources. 

It is easy to observationally test Equation (12) using RM results. 
Figure 6{\it b} plots $Y$ versus $\mathdotM$.  It is very clear that the 
observed data for objects with $\mathdotM<3$ agree well with Equation 
(12). Furthermore, there is a clear saturation of $Y$ for objects with 
$\mathdotM\ge3$ objects. All these results strengthen the conclusions drawn in 
Paper IV.  As in that work, we define an empirical relation 
\begin{equation}
Y=Y_{\rm sat}\min\left[1,\left(\frac{\mathdotM}{\mathdotM_b}\right)^{b}\right],
\end{equation}
where 
\begin{equation}
Y_{\rm sat}=(3.5_{-0.5}^{+0.6})\times 10^4,~~~ 
\mathdotM_b=15.6_{-9.1}^{+22.0},~~~
b=0.27_{-0.04}^{+0.04}.
\end{equation}
In fact, we can get $Y$ by inserting Equation (3) into (7), and find that it 
is in agreement with Equation (13). 
From the saturated$-Y$, we have the maximum value of 
\begin{equation}
    r_{_{\rm H\beta,sat}}=(3.5_{-0.5}^{+0.6}) \times 10^4\,m_7^{-0.29}~~{\rm or}~~ 
    R_{{\rm H\beta,sat}}=(19.9_{-2.8}^{+3.4}) m_7^{0.71}\,{\rm ltd}.
\end{equation}
This result provides a strong constraint on theoretical models of
super-Eddington accretion onto BHs.

\subsection{The Shortened Lags}
The shortened H$\beta$ lags is the strongest distinguishing characteristic 
so far identified between super- and sub-Eddington AGNs. Two factors may lead 
to shortened lags for SEAMBHs.  First, Wang et al. (2014c) showed that, in 
the  Shakura-Sunyaev regime, retrograde accretion onto a BH can lead to 
shorter H$\beta$ lags.  The reason is due to the suppression of ionizing 
photons in retrograde accretion compared with prograde accretion. The 
second factor stems from self-shadowing effects of the inner part of
slim disks (e.g., Li et al. 2010), which efficiently lower the ionizing flux
received by the BLR (Wang et al. 2014c).  When $\mathdotM$ increases, the ratio
of the disk height to disk radius increases due to radiation pressure; 
the radiation field becomes anisotropic (much stronger than the factor of 
$\cos i$) due to the optically thick funnel of the inner part of the slim disk.
In principle, the radiation from a slim disk saturates ($\propto \ln 
\mathdotM$), and the total ionizing luminosity slightly increases with 
accretion rate, but the self-shadowing effects efficiently suppress the 
ionizing flux to the BLR clouds. For face-on disks of type 1 AGNs, observers 
receive the intrinsic luminosity.  If the ionization parameter is constant, 
the ionization front will significantly shrink, and hence the H$\beta$ lag is 
shortened in SEAMBHs compared with sub-Eddington AGNs of the same luminosity. 

The shortened H$\beta$ lag observed in SEAMBHs cannot be caused by retrograde 
accretion.  However, the strong dependence on accretion rate of the deviation 
from the standard lag-luminosity relation implies that the properties of the
ionizing sources are somehow different from those in sub-Eddington AGNs.
According to the standard photoionization theory,
the observed $\rhb\propto L_{5100}^{1/2}$ relation can be explained if
$L_{5100}\propto L_{\rm ion}$ and $Q_c=Un_e\pepsilon$ is constant, where 
$U=L_{\rm ion}/4\pi \rhb^2cn_e\pepsilon$, $L_{\rm ion}$ is the ionizing 
luminosity, $n_e$ is gas density of BLR clouds, and $\pepsilon$ is the 
average energy of the ionizing photons (Bentz et al. 2013). The relation 
$L_{5100}\propto L_{\rm ion}$ holds for sub-Eddington AGNs, and the constancy 
of $Q_c$ is determined by the clouds themselves.  $Q_c$ is not expected to 
vary greatly as a function of Eddington ratio.  Therefore, 
\begin{equation}
\rhb=\frac{L_{\rm ion}^{1/2}}{Q_c}=\calS\rhb^0,
\end{equation}
where the factor $\calS=\left(L_{\rm ion}/L_{\rm ion,0}\right)^{1/2}$ 
describes the anisotropy of the ionizing radiation field, $L_{\rm ion}$ is the 
shadowed ionizing luminosity received by the BLR clouds, and $\rhb^0$ is the 
BLR radius corresponding to $L_{\rm ion,0}$, the unshadowed luminosity. Based 
on the classical model of slim disks, Wang et al. (2014c) showed that, for a 
given accretion rate, $\calS$ strongly depends on the orientation of the 
clouds relative to the disk, and that it range from 1 to a few tens. Therefore,
the reduction of the H$\beta$ lag can, in principle, reach up to a factor of a 
few, even 10, as observed.

Furthermore, the saturated-$Y$ implies that the ionizing luminosity received 
by the BLR clouds gets saturated. The theory of super-Eddington accretion onto 
BHs is still controversial.  Although extensive comparison with models is 
beyond the scope of this paper, we briefly discuss the implications of the 
current observations to the theory.  Two analytical models, which reach 
diametrically extreme opposite conclusions, have been proposed.  Abramowicz et 
al. (1988) suggested a model characterized by fast radial motion with 
sub-Keplerian rotation and strong photon-trapping. Both the shortened lags and 
saturated$-Y$ may be caused by self-shadowing effects and saturated radiation 
from a slim disk.  Both features are expected from the Abramowicz et al. model 
(Wang et al. 2014b). On the other hand, photon-bubble instabilities may govern 
the disk structure and lead to very high radiative efficiency (Gammie 1998).  
If super-Eddington accretion can radiate as much as $L/L_{\rm Edd}\gtrsim 470\,
m_7^{6/5}$ (Equation 14 in Begelman 2002), the disk remains geometrically 
thin. In such an extreme situation, self-shadowing effects are minimal, 
H$\beta$ lags should not be reduced, and $Y-$saturation disappears.

\begin{figure}[!h]
\begin{center}
\includegraphics[angle=0,width=0.48\textwidth]{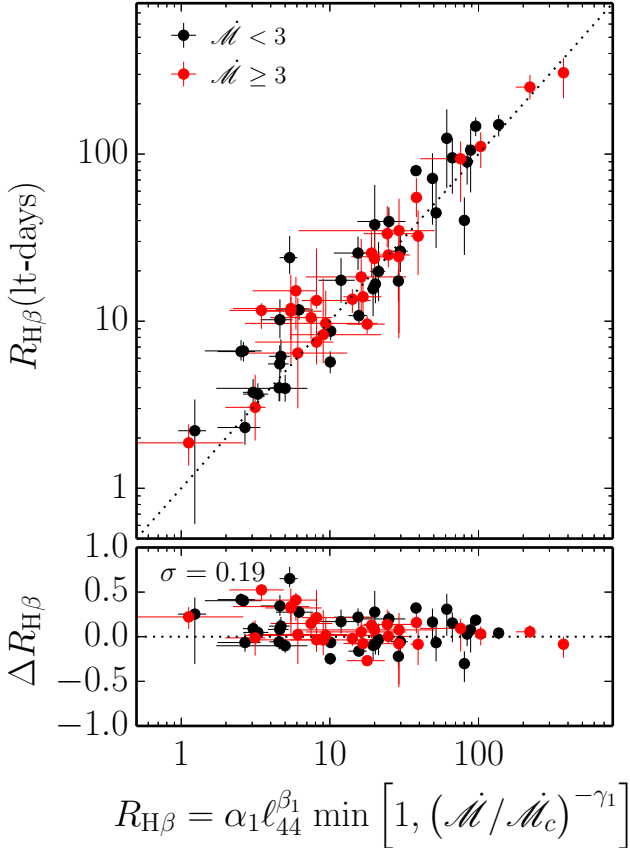}
\end{center}
\vspace{-0.7cm}
\caption{\footnotesize 
The best fit of the new scaling relation for all mapped AGNs. We find that 
$\alpha_1=(29.6_{-2.8}^{+2.7})$\,lt-d, $\beta_1=0.56_{-0.03}^{+0.03}$, 
$\gamma_1=0.52_{-0.16}^{+0.33}$ and $\mathdotM_c=11.19_{-6.22}^{+2.29}$. 
The scatter of the BLR size is greatly reduced to $\sigma=0.19$.
}
\label{r-l}
\end{figure}

Recent numerical simulations that incorporate outflows (e.g., Jiang et al. 
2014) and relativistic jets (Sadowski et al. 2015) also suggest that 
super-Eddington accretion flows can maintain a high radiative efficiency.  
However, most AGNs with high accretion rates are radio-quiet (Ho 2002; Greene 
\& Ho 2006), in apparent contradiction with the numerical simulation 
predictions.  Furthermore, evidence for $Y$-saturation also does not support 
the models with high radiative efficiency.  Recent modifications of the 
classical slim disk model that include photo-trapping appear promising
(e.g., Cao \& Gu 2015; Sadowski et al. 2014), but the situation 
is far from settled.  Whatever the outcome, the results from our observations 
provide crucial empirical constraints on the models.

\subsection{Inclination Effects on $\mathdotM$}
If the BLR is flattened, its inclination angle to the observer will
influence $\bhm$, and hence $\mathdotM$ (see Equation 3).
To zero-order approximation, the observed width of the broad emission lines 
follow
\begin{equation}
\Delta V_{\rm obs}\approx \left[\left(\frac{H_{\rm BLR}}{R}\right)^2+\sin^2i\right]^{1/2}V_{\rm K},
\end{equation}
where $V_{\rm K}$ is the Keplerian velocity and $H_{\rm BLR}$ is the height of the flatten 
BLR (e.g., Collin et al. 2006). For a geometrically thin BLR, $H_{\rm BLR}/R\ll1$, 
$\Delta V_{\rm obs}\approx V_{\rm K}\sin i$, and hence $\mathdotM\propto (\sin i)^{-4}$, 
which is extremely sensitive to the inclination can be severely overestimated for low 
inclinations. On the other hand, many arguments (e.g., Goad \& Korista 2014) support 
$H_{\rm BLR}/R\lesssim 1$, and the inclination angle significantly influences $\bhm$ 
only for $\sin i\gtrsim H_{\rm BLR}/R$. Currently, the values of $H_{\rm BLR}/R$ are 
difficult to estimate, but detailed modelling of RM data suggests $H_{\rm BLR}/R\sim 1$ 
(Li et al. 2013; Pancoast et al. 2014).  If true, this implies that the BLR is not very 
flattened, and hence the inclination angle only has a minimal influence  on $\bhm$ and 
$\mathdotM$.

\subsection{Comparison with Previous Campaigns}
The objects in our SEAMBH sample are very similar to NLS1s.  As previous RM AGN
samples include NLS1s, why have previous studies not noticed that NLS1s deviate
from the $\rblrl$ relation (e.g., Figure 2 in Bentz 2011)?  We believe that the
reason is two-fold.  First, the number of NLS1s included in previous RM
campaigns was quite limited (Denney et al. 2009, 2010; Bentz et al. 2008, 2009;
see summary in Bentz 2011).  The level of optical variability in NLS1s is
generally very low (Klimek et al. 2004), and many previous attempts at RM have
proved to be unsuccessful (e.g., Giannuzzo \& Stirpe 1996; Giannuzzo et al.
1999).  Second, not all NLS1s are necessarily highly accreting.  Our SEAMBH
sample was selected to have high accretion rates (see $\mathdotM$ listed in
Table 7 of Paper IV), generally higher than that of typical NLS1s previously
studied successfully through RM.  As discussed in Wang et al. (2014b) and in
Paper IV, high accretion rates lead to anisotropic ionizing radiation, which
may explain the shortened BLR lags.

\begin{figure*}
\begin{center}
\includegraphics[angle=0,width=0.95\textwidth]{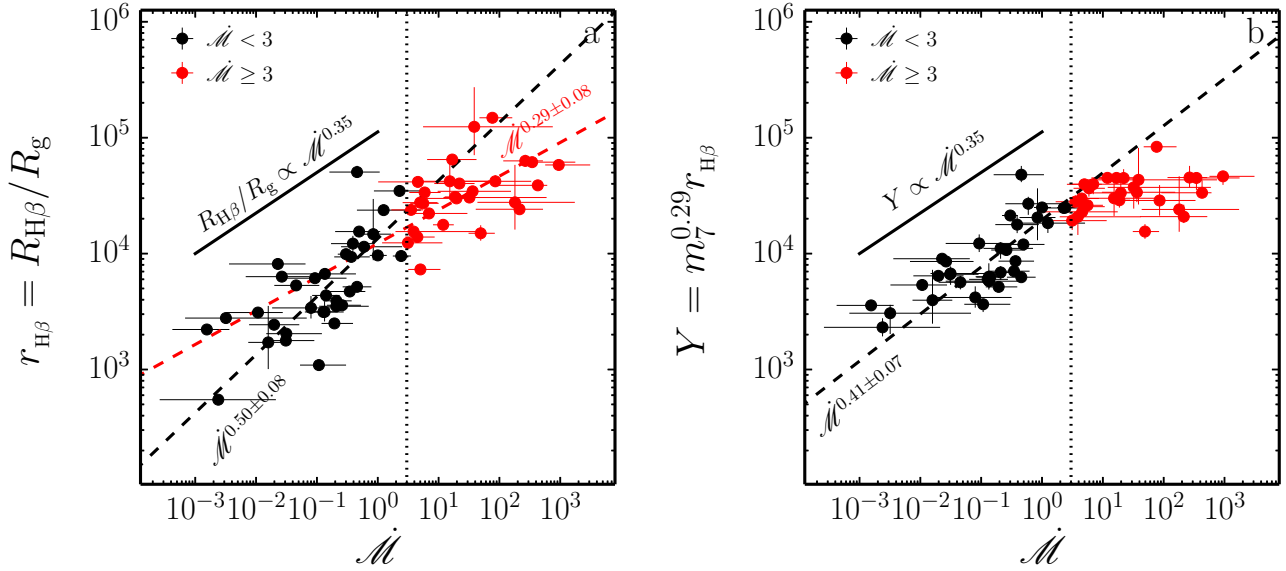}
\end{center}
\vspace{-0.7cm}
\caption{\footnotesize 
The dimensionless BLR radius and $Y$-parameter versus accretion rate 
$\mathdotM$. There is a trend of saturation of $\rrhb$ as shown in panel 
{\it a}, but it is caused by the scatter of BH mass.  
Panel {\it b} shows much a tighter relation and unambiguous saturation of the $Y-$parameter. We 
should point out that NGC 7469 in this plot has been revised compared with Figure 5 in Paper IV, using the 
latest observation from Peterson et al. (2014).
}
\label{r-l}
\end{figure*}

\subsection{SEAMBHs as Standard Candles}
Once its discovery, quasars as the brightest celestial objects in the Universe had been suggested
for cosmology (Sandage 1965; Hoyle \& Burbidge 1966; Longair \& Scheuer 1967; Schmidt 1968; Bahcall
\& Hills 1973; Burbidge \& O'Dell 1973; Baldwin et al. 1978). Unfortunately, the diversity of observed 
quasars made these early attempts elusive. After five decades since its discovery, quasars are 
much well understood: accretion onto supermassive black holes is powering the giant radiation, in 
particular, the BH mass can be reliably measured. Quasars as the most powerful emitters renewed 
interests for cosmology in several independent ways: 1) the normal $\rblrl$ relation (Horn et al. 
2003; Watson et al. 2011; Czerny et al. 2013); 2) the linear relation between BH mass and luminosity 
in super-Eddington quasars (Wang et al.
2013; Paper-II); 3) Eddington AGNs selected by eigenvector 1 (Marziani \& Sulentic 2014); 4) X-ray 
variabilities (La Franca et al. 2014) and 5) $\alpha_{_{\rm OX}}-L_X$ relation (Risaliti \& Lusso 
2015). These parallel methods will be justified for cosmology by their feasibility of experiment 
periods and measurement accuracy. 
 
The strength of SEAMBHs makes its application more convenient for cosmology. Selection of SEAMBHs 
only depends on single epoch spectra through the fundamental plane (Equation 10). BH mass can be
estimated by the new scaling relation (Equation 7). We will apply the scheme outlined by Wang et al. 
(2014a) to the sample of selected SEAMBHs for cosmology in a statistical way (in preparation). 
On the other hand, 
the shortened H$\beta$ lags greatly reduce monitoring periods if SEAMBHs are applied as 
standard candles, in particular, the reduction of lags govern by super-Eddington accretion can 
cancel the cosmological dilltion factor of $(1+z)$. Otherwise, the monitoring periods of 
sub-Eddington AGNs should be extended by the same factor of $(1+z)$ for measurements of H$\beta$ 
lags. Such a campaign of using the normal $\rblrl$ relation for cosmology will last
for a couple of years, even 10 years for bright high$-z$ quasars. Similarly to extension of the 
$\rblrl$ relation to \mgii- and \civ-lines (Vestergaard \& Peterson 2006),
we can extend Equation (7) to \mgii\, and \civ\, lines for the scaling relations
with luminosity as $R_{\rm MgII}(L_{3000},\mathdotM)$ and 
$R_{\rm CIV}(L_{3000},\mathdotM)$, respectively, where $R_{\rm MgII}$ and $R_{\rm CIV}$ are
sizes of the \mgii\, and \civ\, regions, and $L_{3000}$ is the 3000 \AA\, 
luminosity. Such extended relations conveniently allow us to
investigate cosmology by making use of large samples of high$-z$ quasars without
time-consuming RM campaigns. It is urgent for us to make use of kinds of standard candles to
test the growing  evidence for dynamical dark energy (e.g., Zhao et al. 2012; Ade et al. 2015).

\section{Conclusions}
We present the results of the third year of reverberation mapping of super-Eddington
accreting massive black holes (SEAMBHs). H$\beta$ lags of five new SEAMBHs have been detected. 
Similar to the SEAMBH2012 and SEAMBH2013 samples, we find that the SEAMBH2014 objects generally have 
shorter H$\beta$ lags than the normal $\rblrl$ relation, by a factor of a few.
In total, we have detected H$\beta$ lags for 18 SEAMBHs from this project, which have accretion rates 
from $\mathdotM\sim 10$ to $\lesssim 10^3$.  The entire SEAMBH sample allows us to establish 
a new scaling relation for the BLR size, which depends not only on 
luminosity but also on accretion rate. The new relation, applicable over a wide
range of accretion rates from 
$\mathdotM \approx 10^{-3}$ to $10^3$, is given by
$\rhb=\alpha_1\ell_{44}^{\beta_1}\,\min\left[1,\left(\mathdotM/\mathdotM_c\right)^{-\gamma_1}\right]$,
where $\ell_{44}=L_{5100}/10^{44}\,\ergs$ is 5100 \AA\, continuum luminosity, and
coefficients of $\alpha_1=(29.6_{-2.8}^{+2.7})$\,lt-d, $\beta_1=0.56_{-0.03}^{+0.03}$,
$\gamma_1=0.52_{-0.16}^{+0.33}$ and $\mathdotM_c=11.19_{-6.22}^{+2.29}$.

\acknowledgements{ 
We thank an anonymous referee for critical comments that helped to improve the paper.
We acknowledge the support of the staff of the Lijiang 2.4m telescope. 
Funding for the telescope has been provided by CAS and the People's Government of Yunnan 
Province. This research is supported by the Strategic Priority Research Program - The 
Emergence of Cosmological Structures of the Chinese Academy of Sciences, Grant No. XDB09000000, 
by NSFC grants NSFC-11173023, -11133006, -11373024, -11503026,
-11233003 and -11473002, and a NSFC-CAS joint key grant U1431228, and 
by the CAS Key Research Program through KJZD-EW-M06, and by a China-Israel project through NSFC- 11361140347}.

\appendix

\begin{figure*}
\begin{center}
\includegraphics[angle=0,width=0.9\textwidth]{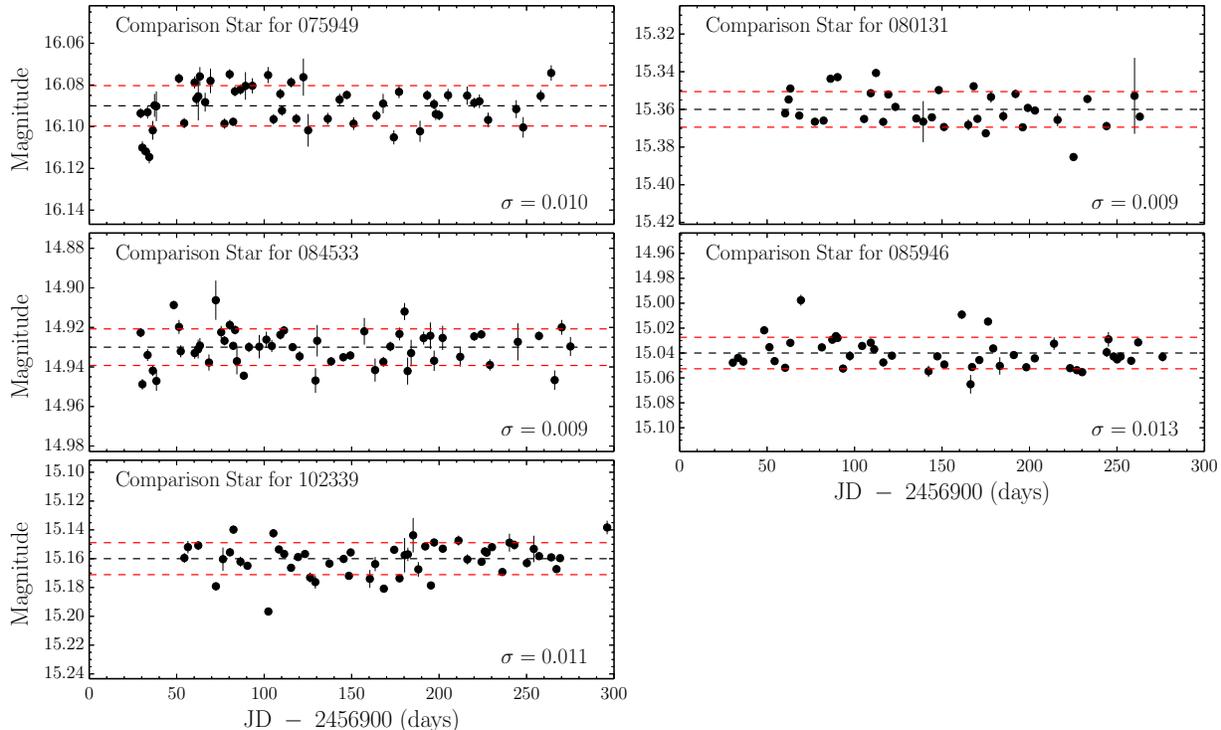}
\end{center}
\vspace{-0.7cm}
\caption{Photometric light curves of comparison stars in the slit.}
\label{r-l}
\end{figure*}

\section{Validity of Equation (3)}
The validity of Equation (3) can be justified for application to SEAMBHs. Solutions of slim disks
are transonic and usually given only by numerical calculations (Abramowicz et al. 1988).
When the accretion rate of the disk is high enough, the complicated structure 
of the disk reduces to a self-similar, analytical form (Wang \& Zhou 1999). Using the 
self-similar solutions (Wang \& Zhou 1999; Wang et al. 1999), we obtained the radius of disk 
region emitting optical (5100  \AA) photons,
\begin{equation}
\frac{R_{5100}}{R_{\rm Sch}}\approx 4.3\times 10^3\,m_7^{-1/2},
\end{equation}
and the photon-trapping radius
\begin{equation}
\frac{R_{\rm trap}}{R_{\rm Sch}}\approx 144\left(\frac{\mathdotM}{10^2}\right).
\end{equation}
We used the blackbody relation $kT_{\rm eff}=hc/\lambda$, where $k$ is the Boltzmann constant,
$T_{\rm eff}$ is the effective temperature of the disk surface,
$h$ is the Planck constant, and $R_{\rm Sch}=3.0\times 10^{12}m_7$\,cm is the Schwartzschild radius. 
Equation (3) holds provided $R_{5100}\gtrsim R_{\rm trap}$, namely 
\begin{equation}
\mathdotM\lesssim 3\times 10^3\,m_7^{-1/2}.
\end{equation}
In this regime, optical radiation is not influenced by photon-trapping effects. We would also like to 
point out that the BH spin only affects emission from the innermost regions of the accretion disk rather 
than the regions emitting 5100  \AA\, photons. In the present campaign, no SEAMBH so far has been found 
to exceed this critical value. Beyond this critical value of accretion rate, optical photons are trapped 
by the accretion flow. We call this the hyper-accretion regime.

Here the cited $10^{-2}$ below Equation (3) is not a strict value of the ADAF threshold 
since it depends on several factors, such as viscosity and outer boundary conditions.
There are a few mapped AGNs with $\mathdotM\lesssim10^{-2}$ (NGC 4151, NGC 5273, 
3C 390.3 and NGC 5548; see Table 7 in Paper IV), but we do not discuss them 
in this paper because they do not influence our conclusions. 

Recently, reprocessing of X-rays (e.g., Frank et al. 2002; Cackett et al. 2007) 
has been found to play an important role in explaining the variability properties of NGC 
5548 (e.g., Fausnaugh et al. 2015). 
The fraction of X-ray emission to the bolometric luminosity strongly anti-correlates with the 
Eddington ratio (see Figure 1 in Wang et al. 2004).  This result is usually interpreted to mean 
that the hot corona becomes weaker with increasing accretion rate, as a result
of more efficiently cooling of the corona
by UV and optical photons from the cold disk. This suggests 
that AGNs with high accretion 
rates will have less reprocessed emission, such that Equation (3) would be more robust in SEAMBHs.

\section{Light curves of comparison stars}
In order to avoid selecting variable stars as comparison stars, we examined their variability. 
To test the invariance of the comparison stars used in our spectroscopic observation, 
we performed differential photometry by comparing them with other stars in the same field.
We typically use six stars for the differential photometry. The light curves 
of the stars are shown in Figure 7. On average, the standard deviations in the light curves
of the comparison stars are 1\%. This guarantees that they can be used as standards
for spectral calibration. 
Figure 7 shows the light curves of the comparison stars for each SEAMBH targets.

\begin{figure*}
\begin{center}
\includegraphics[angle=0,width=0.9\textwidth]{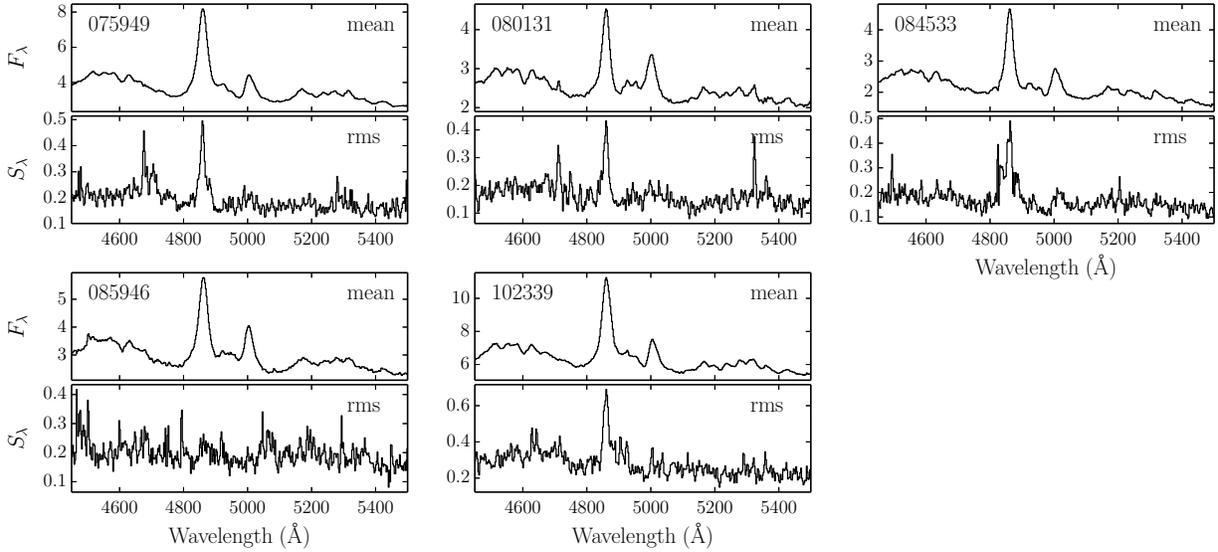}
\end{center}
\caption{\footnotesize 
The averaged and RMS spectra of the SEAMBH2014 sources. $S_{\lambda}$ and $\bar{F}_{\lambda}$ are in units of
$10^{-16}{\rm erg\,s^{-1}\,cm^{-2} \AA^{-1}}$.}
\label{r-l}
\end{figure*}

\section{Averaged and RMS spectra}
The averaged and RMS spectra of the SEAMBH2014 sample are provided in this Appendix. Following the
standard way, we calculated the averaged spectrum as
\begin{equation}
\bar{F}_{\lambda}=\frac{1}{N}\sum_{i=1}^NF_{\lambda}^i,
\end{equation}
and the RMS spectrum as
\begin{equation}
S_{\lambda}=\left[\frac{1}{N}\sum_{i=1}^N\left(F_{\lambda}^i-\bar{F}_{\lambda}\right)^2\right]^{1/2},
\end{equation}
where $F_{\lambda}^i$ is the $i-$th observed spectrum and $N$ is the total number of observed spectra.
They are shown in Figure 8. We note that both the averaged and RMS spectra are
affected by the broadening effects of the $5^{\prime\prime}$-slit on the observed profiles. Using 
the Richards-Lucy iteration, we can correct the observed profiles (averaged and RMS) for 
velocity-resolved mapping, which will be carried out in a separate paper (Du et al. 2015c).

\section{Notes on individual objects}
We briefly remark on individual objects for which H$\beta$ lags have been successfully measured.  We 
failed in getting lag measurements for the other five objects because either their flux variations are 
very small or the data sampling rate was inadequate.

{\it J075949}: The detected H$\beta$ lag arises from two major peaks in the
light curves. 

{\it J080131}: The monitoring observations during $2013-2014$ did not yield a well-determined  H$\beta$ 
lag because of the lack of H$\beta$ response to the second continuum flare (see Paper IV).  During 
$2013-2014$, the first reverberation of H$\beta$, 
which can be clearly seen during the first 70 days of the light curves, yields a very significant lag, 
as shown in the CCF with a rest-frame centroid lag of $11.5_{-3.6}^{+8.4}$ days (with a very high 
coefficient of $r_{\rm max}=0.81$). We monitored this object again in this observing 
season (Figure 1). We successfully measure $\tauhb=11.2_{-9.8}^{+14.8}$ days, consistent with last 
season's result.

{\it J084533}: Its continuum slightly decreased before being monitored for $\sim 70$ days, and
steadily increased until $\sim 200$ days and then decreased again. Although
the CCF has a very flat peak close to $\sim 0.9$, Monte Carlo simulations show that H$\beta$ 
lag is around 20 days, which arises from the two peaks in the H$\beta$ and $r^{\prime}$-band 
light curves.  

{\it J085946}: The CCF peaks near 0.6, which results from the two major dips in
the H$\beta$ and $r^{\prime}$-band light curves. There are two peaks
    with roughly the same correlation coefficients around $\sim$20 days and
    $\sim$70 days in the observed frame, respectively. Considering the relatively poor data 
    quality of this object, it is difficult to distinguish which is the true response.
    The centroid lag represents the average of
    these two peaks (responses), and its uncertainties cover the distribution
(Figure 1) obtained in FR/RSS method. So, we use it in the analysis of main
text.

{\it J102339}: The detected lag is from the dip feature around $\sim 150$ days in light curves.

\section{Description of CCCD}
For multiple-peaked CCFs with similar correlation coefficients, it is ambiguous
as to which peak should be used to calculate the final lag. In such cases, we
use the CCCD to determine the lag. However,  there are two approaches to
calculate the centroid time lag in the CCF, as illustrated in Figure 9. 
\begin{itemize}
\item Approach 1 calculates the centroid using all peaks above some criterion,
such as $0.8r_{\rm max}$.
\item Approach 2 only uses the highest peak. 
\end{itemize}
In Approach 1, the CCCD tends to be smoother than the CCPD (see Figure 1),
whereas in in Approach 2 the CCCD and CCPD always have a similar distribution.
We adopt Approach 1 in our analysis. If the CCF has two or even three peaks,
the two approaches give different centroid lags.  However, when the quality of
the data is high and the CCF is unimodal, the two approaches yield the same
results. It should be pointed out that Approach 2 is often employed in the
literature.

\begin{figure*}
    \centering
    \includegraphics[width=0.80\textwidth]{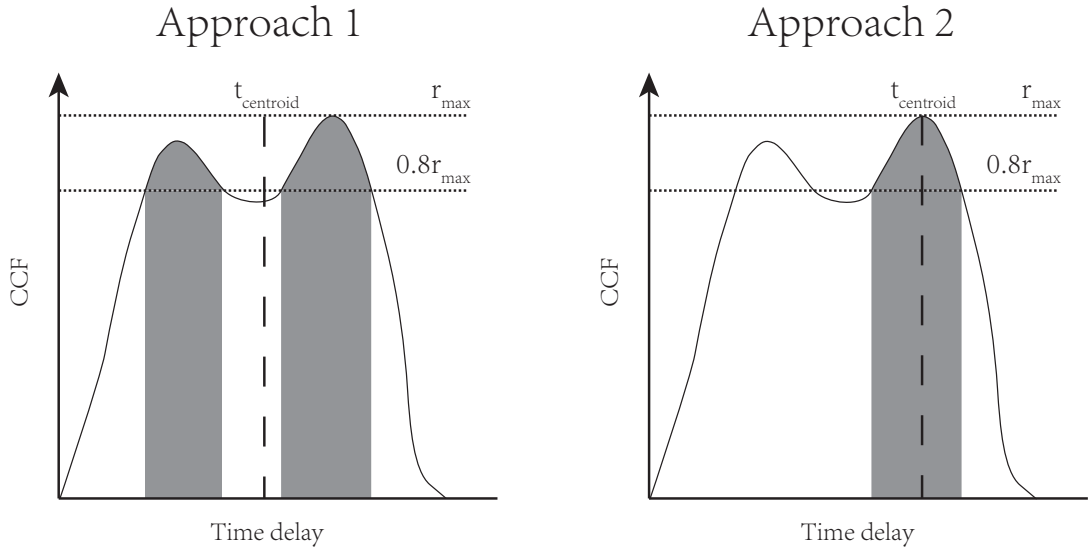}
    \caption{\footnotesize Two approaches to calculate centroid time lag. We adopted 
    Approach 1 in our analysis. 
}\label{fig:approaches}
\end{figure*}

\end{document}